\newif\ifdraft
\drafttrue
\documentclass [
  smallextended
]{svjour3}

\usepackage [utf8] {inputenc}
\usepackage [T1] {fontenc}
\usepackage [super] {nth}

\usepackage [
  dvipsnames,
  table
] {xcolor}

\usepackage {
  amssymb,
  booktabs,
  csquotes,
  listings,
  lst-yaml,
  microtype,
  natbib,
  pifont,
  setspace,
  subcaption,
  tabularx,
  textcomp,
  tikz,
  url,
  xspace,
}

\usepackage [normalem] {ulem}

\usepackage [scaled = 0.76] {beramono}
\usepackage [most] {tcolorbox}

\usepackage[
  pdftex,
  colorlinks = true,
  bookmarks  = false,
  linkcolor  = RoyalBlue,
  citecolor  = RoyalBlue,
  urlcolor   = RoyalBlue,
  bookmarks
]{hyperref}

\usepackage[
  linecolor = RoyalBlue,
  bordercolor = RoyalBlue,
  backgroundcolor = RoyalBlue,
  textsize = scriptsize,
  textwidth = 19mm,
]{todonotes}

\makeatletter
\def\cl@chapter{\@elt {theorem}}
\makeatother

\usepackage[capitalize]{cleveref}
\usepackage{crossreftools}
\crefname{figure}{Fig.}{figures}
\Crefname{figure}{Figure}{Figures}
\crefname{table}{Tbl.}{tables}
\Crefname{table}{Table}{tables}
\crefname{section}{Sect.}{sections}
\Crefname{section}{Section}{Sections}

\newcolumntype L {>{\raggedright\arraybackslash}X}
\newcolumntype s {>{\raggedright\arraybackslash}m{2.8cm}}

\newcommand \repourl
  {\url {https://github.com/robust-rosin/robust}\xspace}

\newcommand \subject [1]    {#1\xspace}
\newcommand \robust         {\texttt {ROBUST}}
\newcommand \kobuki         {\subject {Kobuki}}
\newcommand \turtlebot      {\subject {TurtleBot}}
\newcommand \mavros         {\subject {MavRos}}
\newcommand \universalrobot {\subject {Universal\,Robot}}
\newcommand \motoman        {\subject {Motoman}}
\newcommand \careobot       {\subject {Care-O-bot}}
\newcommand \rosgeometry    {\subject {geometry2}\xspace}

\newcommand \bugg [3] {%
  {\href {https://github.com/robust-rosin/robust/tree/master/#2/#3/#3.bug}%
  {\texttt {\color {black}\textls [-20] {#1\kern 0.15em\clap :\kern .15em\allowbreak #3}\xspace}}}%
}

\newcommand{\examples}[1]{{\color{black}\texttt[\kern-0.20ex#1\kern-0.23ex\texttt]}}

\newcommand {\bugkobuki} [1]
  {\bugg{ko\-bu\-ki}{kobuki}{#1}}
\newcommand {\bugturtlebot} [1]
  {\bugg{tur\-tle\-bot}{turtlebot}{#1}}
\newcommand {\bugmavros} [1]
  {\bugg{mav\-ros}{mavros}{#1}}
\newcommand {\buguniversalrobot} [1]
  {\bugg{uni\-ver\-sal\_ro\-bot}{universal\_robot}{#1}}
\newcommand {\bugmotoman} [1]
  {\bugg{mo\-to\-man}{motoman}{#1}}
\newcommand {\bugcareobot} [1]
  {\bugg{careobot}{care-o-bot}{#1}}
\newcommand {\bugconfidential} [1]
  {\bugg{con\-fi\-den\-tial}{confidential}{#1}}
\newcommand {\buggeometry} [1]
  {\bugg{ge\-o\-me\-try}{geometry2}{#1}}
\newcommand {\bugroscomm} [1]
  {\bugg{ros\-comm}{ros\_comm}{#1}}
\newcommand {\bugother} [1]
  {\bugg{o\-ther}{other}{#1}}

\newcommand {\numbugs} {221\xspace}
\newcommand {\numsystems} {seven\xspace}

\newcommand {\numbuildtimebugs} {63\xspace}
\newcommand {\numstartuptimebugs} {38\xspace}
\newcommand {\numruntimebugs} {118\xspace}
\newcommand {\numruntimeandstartuptimebugs} {156\xspace}

\newcommand{\numfixedbugswithtests}{15\xspace}

\newcommand {\numbdo} {77\xspace}
\newcommand {\numbugslossfun} {42\xspace}
\newcommand {\numbugsunresponsive} {26\xspace}
\newcommand {\numbugssysperf} {7\xspace}
\newcommand {\numbugssysperftext} {seven\xspace}
\newcommand {\numbugssysbehaviour} {44\xspace}
\newcommand {\numbugsmonitor} {11\xspace}
\newcommand {\numbugssysnone} {106\xspace}
\newcommand {\numbugssys} {115\xspace}

\newcommand {\numbugsysnonebutnotbuild} {43\xspace}

\newcommand {\numbugsswbuild} {54\xspace}
\newcommand {\numbugsswui} {20\xspace}
\newcommand {\numbugsswperf} {2\xspace}
\newcommand {\numbugsswcrash} {40\xspace}
\newcommand {\numbugsswliveness} {15\xspace}
\newcommand {\numbugsswbehavioral} {68\xspace}
\newcommand {\numbugsswnetwork} {7\xspace}

\newcommand*\emptycirc[1][.8ex]{\tikz\draw[thick] (0,0) circle (#1);}
\newcommand*\halfcirc[1][.8ex]{%
  \begin{tikzpicture}
  \draw[fill] (0,0)-- (90:#1) arc (90:270:#1) -- cycle ;
  \draw[thick] (0,0) circle (#1);
  \end{tikzpicture}}
\newcommand*\fullcirc[1][.8ex]{\tikz\fill (0,0) circle (#1);}

\makeatletter
\newcommand{\bianca}{\renewcommand\NAT@open{[}\renewcommand\NAT@close{]}}
\makeatother
\newcommand*\sqcitep[1]{{\bianca\citep{#1}}}

\def\CC{{C\nolinebreak[4]\hspace{-.05em}\raisebox{.35ex}{\scriptsize\textbf ++}}\xspace}

\newlength \HEIGHT

\journalname{Empirical Software Engineering}

\begin{document}

\title{ROBUST:\ \numbugs\ Bugs in the Robot Operating System.}

\author{Christopher S. Timperley \and  Gijs van der Hoorn \and Andr\'e Santos \and Harshavardhan Deshpande \and Andrzej Wąsowski}

\institute{Christopher S.\ Timperley\at
              School of Computer Science, Carnegie Mellon University, USA \\
              \email{ctimperley@cmu.edu}%  \\
           \and
           Gijs van der Hoorn \at
           Department of Cognitive Robotics, faculty of Mechanical, Maritime and Materials Engineering, Delft University of Technology, the Netherlands\\
           \email{g.a.vanderhoorn@tudelft.nl}
           \and
           Andr\'e Santos \at
              VORTEX CoLab, Portugal\\
              \email{andre.santos@vortex-colab.com}
           \and
           Harshavardhan Deshpande \at
              Fraunhofer Institute for Manufacturing Engineering and Automation IPA, Germany\\
              \email{harshavardhan.deshpande@ipa.fraunhofer.de}
           \and
           Andrzej Wąsowski \at
           Department of Computer Science, IT University of Copenhagen, Denmark\\
           \email{wasowski@itu.dk}
}

\date{Received: date / Accepted: date}

\maketitle

\begin{abstract}
As robotic systems such as autonomous cars and delivery drones assume greater roles and responsibilities within society, the likelihood and impact of catastrophic software failure within those systems is increased.
To aid researchers in the development of new methods to measure and assure the safety and quality of robotics software, we systematically curated a dataset of 221 bugs across 7 popular and diverse software systems implemented via the Robot Operating System (ROS).
We produce historically accurate recreations of each of the 221 defective software versions in the form of Docker images, and use a grounded theory approach to examine and categorize their corresponding faults, failures, and fixes.
Finally, we reflect on the implications of our findings and outline future research directions for the community.

\keywords{robotics \and software bugs \and dataset \and Robot Operating System \and ROS \and BugZoo}
\end{abstract}

\section*{Declarations}

\begin{description}
  \item[\textbf{Funding:}]
    This work was partially supported by the ROSIN project under the European Union’s Horizon 2020 research and innovation programme, grant agreement No. 73228, and by DARPA (\#A8750-16-2-0042) and AFRL (\#FA8750-15-2-0075).  The authors are grateful for their support.  Any opinions, findings, or recommendations expressed are those of the authors and do not necessarily reflect those of the US Government or the European Union.
  \item[\textbf{Conflicts of interest:}] The authors have no competing interests to declare that are relevant to the content of this article.
  \item[\textbf{Availability of data and material:}]
    The ROBUST dataset together with the Jupyter notebook used to produce figures shown in this paper, is publicly available at the following location: \repourl.
  \item[\textbf{Code availability}:] Not applicable.
\end{description}

\section{Introduction}\label{sec:intro}

From assembling and manufacturing goods to driving us from place to place, robotic systems constitute an increasingly large part of the computing ecosystem.
However, these systems, and the software that controls them, present new opportunities for cyberattacks and catastrophic failures with the potential for enormous economic and human damage\,\citep{verge-737max,space-exomars,ieee-nissan-airbag,nbc-uber-tempe}.
To fully realize the benefits of robotic systems, we need effective quality assurance (QA) techniques for robotics software that allow developers to build advanced applications without compromise to safety. In order to catalyze the development of these QA techniques for robotics, it is important that we better understand the nature of bugs within robotics software.
By better understanding the nature of software bugs in robotic systems, we can identify key challenges and promising avenues of research.
To that end, we have endeavoured  to paint a detailed picture of challenges in the largest software ecosystem for robotics, the Robot Operating System (ROS).
We take the perspective of bugs, previously documented and, mostly, fixed in the open source repositories of ROS code.

ROS, colloquially known as the \enquote{Linux of Robotics,} is a highly modular and distributed open-source platform for building robotics systems with a rich ecosystem of thousands of reusable software packages~\citep{linux-of-robotics,Kolak20}. ROS runs on top of Linux typically.
It is widely used in teaching, research and engineering, and attracts major industrial players including Amazon, Intel, and Microsoft:
The ROS-Industrial Consortium counts close to a hundred organizations including Bosch, Siemens, and Boeing~\citep{rosin-members}.
There is a growing belief that a shared open source platform will allow the industry to exploit the economy of scope in robotics.  High-quality hardware drivers, control modules, AI components, and development tools shall benefit the entire ecosystem while the cost of building them is carried by multiple organizations.
Such benefits shall also be extended to testing and quality assurance tools and methods, which are of paramount importance for professional development of software in industry.
This paper, devoted to the software quality challenges in the ROS ecosystem, seeks to identify opportunities for research and development that will benefit software development for robotics in general and for ROS in particular.

We report the results of a collaboration between academic and industrial partners to document, reproduce, and understand software bugs that occur in ROS software.
Our method is artifact-driven:  we create a data set of ROS bugs and then study it. Thus the main outcome of the work is the artifact.
We systematically identified \numbugs bugs across \numsystems popular and diverse ROS subject systems, representing a variety of domains and layers of the ROS application stack, by studying their respective version control histories and artifacts. We methodically studied each bug to produce a structured forensic description of its causes, symptoms, and fix, amongst other details.
We then used a grounded theory to categorize and understand the nature of faults and failures within ROS systems. To bolster research into effective quality assurance techniques, we package each bug into a historically accurate Docker container image, allowing it to be accurately studied by others for a variety of purposes (e.g., program repair, testing, debugging).

Our analysis of the ROBUST dataset demonstrates the inherent difficulties of building high-quality reusable robotics software components and the unintended consequences of framework design tradeoffs, and identifies opportunities for the development of new languages and analysis tools to mitigate these difficulties.
We find that ROS developers make similar kinds of mistakes as other developers, and that certain bugs within the dataset could have been detected by existing tools and practices (e.g., type checking, fuzzing, continuous integration).
In line with previous research~\citep{AfzalICST20}, we find that very few bugs are accompanied by regression tests, testing is generally lacking across each of the studied systems, and, consequently, developers tend to rely on manual testing efforts.
Finally, we observe that, while failures typically span across multiple components, bug fixes are comparatively simple and confined to a small number of lines within a single file.

The main contributions of this paper include:

\begin{itemize}

  \item A dataset of ROS bugs, ROBUST, containing
  \numbugs bugs across \numsystems popular ROS systems.
  ROBUST is open source and free to use, and can be found at: \url{https://github.io/robust-rosin/robust},

  \item An analysis of the faults and failures of bugs represented within the ROBUST dataset
    (\Cref{sec:method,sec:analysis}),

  \item A discussion of implications for practitioners and the research community (\Cref{sec:findings}),

  \item A method and a toolchain for building a bug repository and accurately replicating bugs in ROS systems and similar ecosystems (\Cref{sec:infrastructure}.)

\end{itemize}

\noindent
\Cref{sec:subjects} provides a basic characterization of ROS as a subject of study and discusses the selection of ROS systems to harvest the bugs from.

Our intended audience are software engineering researchers that build new tools and methods for robotics software developers.
The \robust\ dataset aims to lower the barrier of entry into research on software engineering for robotics, supporting, among others, further work on testing, fuzzing, architectural analysis techniques, verification, program repair, etc.
For example, a \emph{fuzzing tool} for robotics would have to work with multiple executables, multiple input streams, multiple programming languages, and network communication.
\robust\ bugs that manifest themselves in crashes can be used to test effectiveness of such a fuzzer as each of them can be easily re-established in a Docker container.
Building a cross-language \emph{program-repair tool} for Python and \CC requires understanding any cross-language bugs in such systems and of the API binding mechanisms in these languages.
\robust\ can save a lot of time in this process, as it contains more than twenty bugs on the Python/\CC boundary that are ready to be reproduced and used in the tool design, evaluation, or regression testing.
\robust\ can also support building tools for \emph{domain-specific languages} as it contains more than a hundred documented bugs that have been repaired by developers in files written in domain specific languages.
These are just three examples of applications of \robust\ as a research tool.
We genuinely hope that it can also help in work on build systems, software configuration, concurrency, and evolution.
\looseness -1

\section {Subject Ecosystem and Subject Systems}%
\label{sec:subjects}

\paragraph {The Subject Platform: Robot Operating System.}

The Robot Operating System originates from Stanford University\,\citep{Quigley09}. In 2007, the project had transferred to a robotics start-up, Willow Garage, which in turn founded the Open Source Robotics Foundation, a non-profit dedicated to promoting open source robotics and the present steward of the community. ROS started as a communication middleware allowing developers to create robot systems using established architectures for distributed systems. It has quickly expanded with numerous tools, hardware drivers, and software modules for robotics-specific tasks such as planning, navigation, and perception. Today, the core modules are still developed by a tight group of developers, however, the majority of components in the vast ecosystem are maintained by a large community of companies and research groups in a highly decentralized manner. Due to its size and proliferation, ROS is presently the most representative robot software development framework. \looseness -1

ROS software is organized in \emph{packages}, the basic build and release units. Packages can be added to an official index and released in \emph{distributions}, which are updated following a well-defined life cycle pattern similar to that of Ubuntu Linux. For instance, ROS Indigo Igloo, released in May 2014, was a \emph{Long Term Support} (LTS) distribution, targeting Linux Ubuntu Saucy (13.10) and Linux Ubuntu Trusty (14.04 LTS). Support for Indigo ended in April 2019, at the same time that Ubuntu Trusty reached its end-of-life. Current distributions have already surpassed the mark of 4,000 packages. Besides these, there are many community packages that are not part of the official distributions.

A ROS-based system follows a distributed architecture, with independent runtime resources connecting to each other. Every resource is named, using a hierarchical naming structure. Names can be \emph{remapped} during system initialization without changing source code. This mechanisms allows developers to compose systems from different packages without modifying them.

\begin{figure}[t!]% [__ fig:subject-photos
\centering
\begin{subfigure}[b]{.35 \textwidth}
    \centering
    \hbox{\strut\hspace{15mm}\includegraphics[
      height = 24mm,
      clip,
      trim = 0mm 1mm 0mm 0mm
    ]{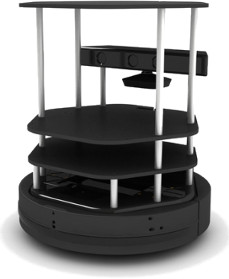}\hspace{15mm}\strut}
    \caption{\turtlebot with a \kobuki base}
    \label{fig:turtlebot2}
\end{subfigure}
\hfill
\begin{subfigure}[b]{0.27 \textwidth}
    \centering
    \includegraphics[
      height = 30mm
    ]{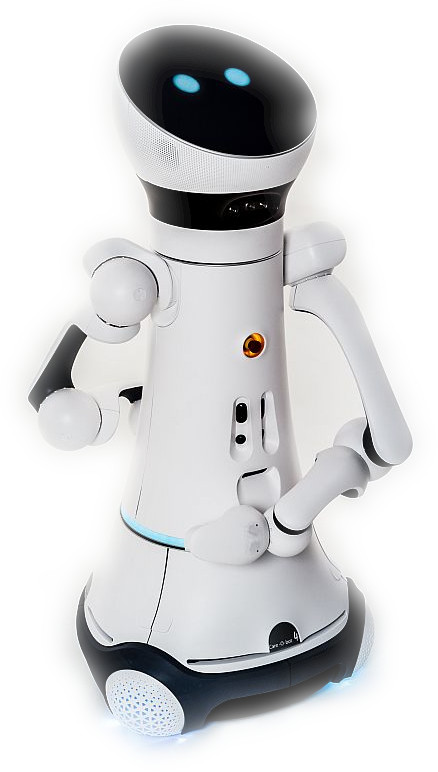}
    \caption{\careobot~4}
    \label{fig:cob}
\end{subfigure}
\hfill
\begin{subfigure}[b]{0.28\textwidth}
    \centering
    \includegraphics[
      height = 25mm
    ]{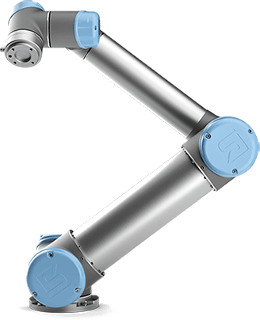}
    \caption{\universalrobot}
    \label{fig:ur}
\end{subfigure}

\begin{subfigure}[b]{0.19\textwidth}
    \centering
    \hbox{\strut\hspace{3mm}\includegraphics[
      height = 27mm,
    ]{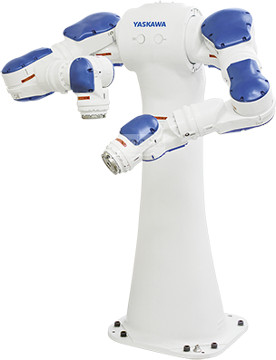}\hspace{8mm}\strut}
    \caption{\motoman}
    \label{fig:motoman}
\end{subfigure}
\hfill
\begin{subfigure}[b]{0.37\textwidth}
    \centering
    \includegraphics[
      height = 21 mm
    ]{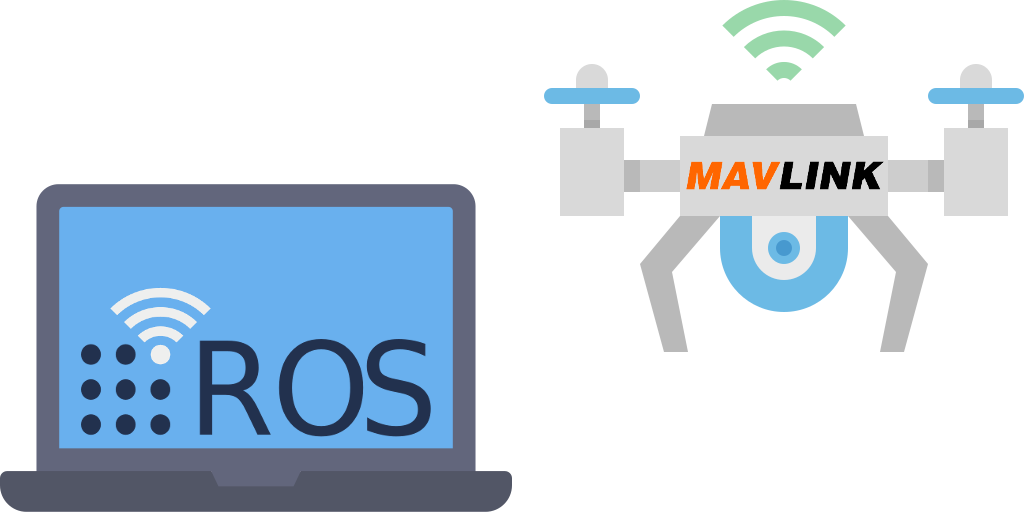}
    \caption{\mavros{$^\ast$}}
    \label{fig:mavros}
\end{subfigure}
\hfill
\begin{subfigure}[b]{0.34 \textwidth}
    \centering
    \includegraphics[
      height = 27mm
    ]{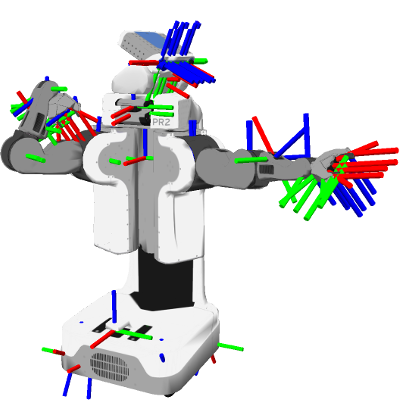}
    \caption{Robot model with \texttt{tf2} frames}
    \label{fig:tf2}
\end{subfigure}
\caption{Subject systems, not to scale. $^\ast$Images: \href{https://www.flaticon.com/authors/dinosoftlabs}{DinosoftLabs}, \href{https://www.flaticon.com/authors/photo3idea-studio}{photo3idea\_studio}.}%
\label{fig:subject-photos}
\end{figure}% __]

ROS supports four types of resources: \emph{parameters}, \emph{nodes}, \emph{topics}, and \emph{services}. Parameters are variables holding shared data, stored in a central Parameter Server. Nodes are programs that consume and produce data, and communicate with each other via message passing channels---asynchronous topics and synchronous services. Topics and services are both typed. All nodes are expected to obey types handled by services and topics, which is enforced at runtime. Users can define custom messages adding to the available predefined types representing primitive data, laser scans, 3D poses, etc.

Topics are the most common message-passing mechanism. They follow an asyn\-chronous publisher-subscriber model, with many-to-many connections. Publishers can send messages at any time, regardless of the number of active subscribers, and subscribers are notified via a callback function whenever a new message arrives. Services implement synchronous one-to-one communication, using remote procedure calls. The node that provides a service is called the server, and any nodes using the service are called clients. Messages are exchanged in request-response transactions. Clients block while waiting for a response. Services shall be used for fast tasks, such as querying the current state of a node.
\looseness -1

Nodes should be specific and modular, rather than large monolithic components. A single robot consists of many nodes, each accomplishing a task, such as localization, navigation, or perception. Each sensor and actuator might have its individual node, too. Nodes are implemented using a ROS client library. Many programming languages are supported but C++ and Python are most used.
\looseness -1

\begin{table}[p!]% [__ tab:subject-descriptions

  \small

  \caption{Descriptions of subject systems. See also \Cref{tab:subjects} and \Cref{fig:subject-photos}.}%
  \label{tab:subject-descriptions}

  \begin{tabularx}{\textwidth}{
    @{}
    X
    X
    @{}
  }

    \textbf{\turtlebot} is an iconic ROS robot (\cref{fig:turtlebot2}),  a low-cost entry-level personal robot kit with open-source software built with the original authors of ROS.  The software includes example applications for automatic docking, charging, navigation, and dynamic leader-follower behaviors, written mostly in Python.
    \looseness = -1

    & \textbf{\kobuki} is the mobile base on top of which \turtlebot built, the black disk at the bottom of the robot in \cref{fig:turtlebot2}.  The \kobuki software repositories provide low-level packages that integrate the base hardware (servomotors, power systems) with ROS, as well as a safety controller and a velocity multiplexer, mostly written in C++.
    \looseness = -1
    \\

    \textbf{\mavros} implements a bridge between ROS and the MAVLink protocol for communicating with the autopilot of unmanned vehicles (air, ground, and water).  The MAVROS repository provides tools and plugins that allow almost transparent bridging between a MAVLink-enabled autopilot, such as ArduPilot, and a ROS application, making as much of the data gathered and processed by these systems available to the ROS node graph.  The repository contains only the bridging nodes (in C++/Python), and explicitly leaves the modelling of vehicle geometry, kinematics, and dynamics to other packages.
    \looseness = -1

    & \textbf{\universalrobot} and \textbf{\motoman} provide packages for the integration of industrial robots with ROS  (\cref{fig:ur,fig:motoman}, both robots are industrial manipulators).  Packages, implemented mostly in C++, Python, and C, include low-level interfaces to the motion controllers of the robot, sensors and I/O interfaces, as well as higher-level declarative description packages that provide information on robot geometry, kinematics, dynamics, and motion planning configurations.  Both repositories also include programs that are to be executed on the industrial robot controller itself, and which will collaborate with their ROS counterparts.
    \looseness = -1
    \\

    \textbf{\careobot} is an autonomous service robot, created by Fraunhofer IPA and commercialized by Mojin Robotics (\cref{fig:cob}).  Nearly all its behavior is implemented as ROS nodes (a significant part open-source) including low-level interfaces to motor controllers and sensors, human-machine interfaces (speech, face, emotion and gesture recognition), collision-free path planning, manipulation planners, object recognition, and localization.  Packages for simulation of the entire system are also included, as well descriptions of its geometry, kinematics, and dynamics.  Most of the code is written in C++ and Python.  A rather large amount of XML is used mostly to define the many configurations enabled off-the-shelf.

    & \textbf{Geometry2}, also known as \texttt{tf2}, offers a set of services for keeping track of multiple coordinate frames over time and for transforming points and other geometric objects between frames.  The \rosgeometry libraries are implemented in C++ and interfaced to Python. They can be used to determine the global position of the robot relative to the world or the position of a gripper relative to the robot base.  Applications range from localization and visualization to mapping and multi-robot cooperation.  \Cref{fig:tf2} shows a 3D robot model annotated with coordinate frames tracked by \texttt{tf2} on the robot's base, head, arms and grippers (visualized as three colored axes).
    \looseness = -1

  \end{tabularx}

  \caption{Descriptive statistics for the subject systems.
  The second and third columns show the number of identified bugs and issues that were studied, respectively.
  The last four columns show the number of lines in main languages, counted with \textsc{cloc} v1.60.}%
  \label{tab:subjects}\label{tab:issues}

  \medskip

  \begin{tabularx}{\textwidth}{
    @{}
    >{\bfseries}X
                r
                r
                X
                r
                r
                r
                r@{}
    }
    \textbf{Subject} & \textbf{\# Bugs} & \textbf{\# Issues} & \textbf{Category} & \textbf{C++} & \textbf{C} & \textbf{Python} & \textbf{XML}
    \\[1.5mm]

    \kobuki          & 57 & 325 & application   & 23,555 & 18,073 & 4,207 &  2,325 \\
    \turtlebot       & 11 & 170 & application   &    799 &     42 & 4,438 &  1,129 \\
    \strut\rlap{\careobot} & 11 & 182 & application   & 31,084 &  9,430 & 9,248 & 23,814 \\
    \strut\rlap{\universalrobot} & 25 & 158 & driver        &  1,071 &    331 & 1,741 &    738 \\
    \motoman         & 22 & 78  & driver        &  4,129 &  5,337 &     0 &  1,272 \\
    \mavros          & 40 & 623 & middleware    & 12,807 &  1,611 & 1,013 &    330 \\
    \rosgeometry     & 42 & 264 & library       &  6,267 &  4,311 & 1,074 &    273
    \\[1.5mm]

    \textbf{Total}  & \textbf{208} & \llap{\textbf{1800}} & & \llap{\textbf{79,712}} & \llap{\textbf{39,135}} & \llap{\textbf{21,721}} & \llap{\textbf{29,881}}
  \end{tabularx}

  \vspace{-1mm}

\end{table}% __]

\paragraph {Subject Systems.}

Conducting a meaningful analysis of software bugs requires considerable domain expertise. Therefore, we deliberately set out to examine a diverse set of subject systems for which we had expertise in our research team.
In total, we examined \numsystems\ qualitatively different subject systems consisting of three robotics systems, two robot drivers, and two very different specialized libraries.
\Cref{fig:subject-photos} shows photographs and visualizations of the subject systems, \Cref{tab:subject-descriptions} summarizes their key descriptive properties, and \Cref{tab:subjects} provides a quantitative characterization in the last four columns.

\section {Method}%

\label {sec:method}

\begin {figure} [t!]% [__ fig:github-issue

  \begin {center}

    \includegraphics [
      width = .95 \textwidth,
      clip,
      trim = 5.5mm 0mm 3mm 0mm
    ] {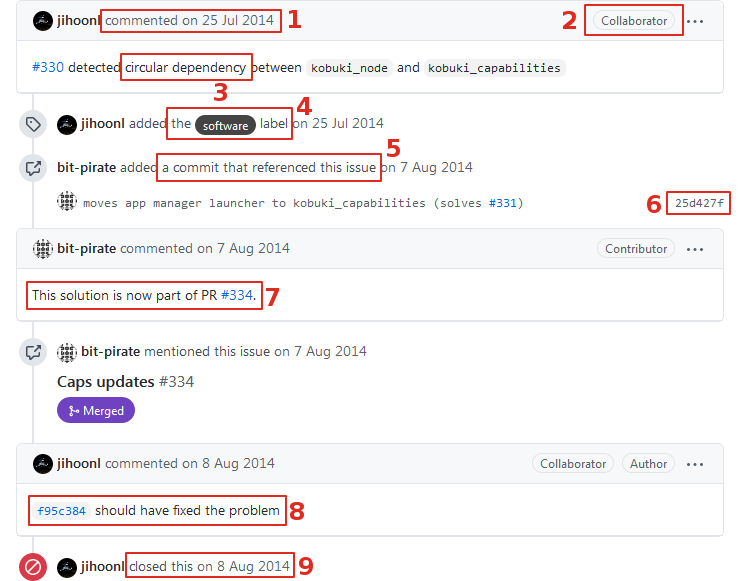}

  \end {center}

  \vspace {-2mm}

  \caption {Issue 331 in \kobuki, \url {https://github.com/yujinrobot/kobuki/issues/331}}%
  \label {fig:github-issue}

\end {figure}% __]

\paragraph {Subject Bugs and Data Gathering.}

To identify historical bugs in each subject system, we examined its issue tracker, pull requests, and commit history. For an initial screening, we prioritized issues labeled as a \emph{Bug} (or similar) and commit messages including keywords such as \emph{fix}. Issues clearly unrelated to bugs by their title or labels were discarded. All non-obvious issues and pull requests were inspected in consensus meetings to determine whether they describe additional bugs. In the meetings, we have asked whether the problem discussed is a result of a deliberate prior design decision, or whether it is a result of an omission, a mistake, a change in another system, etc.  In any case, whenever developers used the term bug, error, or mistake in the discussion, we assumed a bug is being discussed.
Bad smells and style issues were classified as not-bugs.
In total, we identified \numbugs\ issues and pull requests that qualified as bugs across the subject systems.
\looseness -1

\Cref {fig:github-issue} exemplifies the data available about the bugs.
The issue \emph {creation date}\,(1) determines the versions of ROS and other dependencies that might have been used by the reporter.
The \emph{community status of the reporter}\,(2) distinguishes between issues found internally and by the downstream users.
The \emph {problem description}\,(3) is the key source regarding whether an issue is a bug and what is its nature.
The \emph{labels}\,(4) provide a diversity of information.
Here the issue has been labelled as software-related, which makes it a potential \emph{software} bug report.
The existence of \emph{commits}\,(5, 8) referencing the issue show that it is either fixed or being worked on.
Inspecting the \emph{commit}\,(6, 8) we can understand the bug from the perspective of its fix.
The \emph{referencing pull requests}\,(7) provide similar context.
If the bug is fixed, we note down the \emph{closing date}\,(9) of the issue. This allows to estimate how long it took to resolve the problem (14 days here).
\looseness -1

\begin{figure}[t!]% [__ fig:github-commit

  \begin{center}

    \includegraphics[
      width = .89 \textwidth,
      clip,
      trim = 10.5mm 0mm 11mm 0mm
    ]{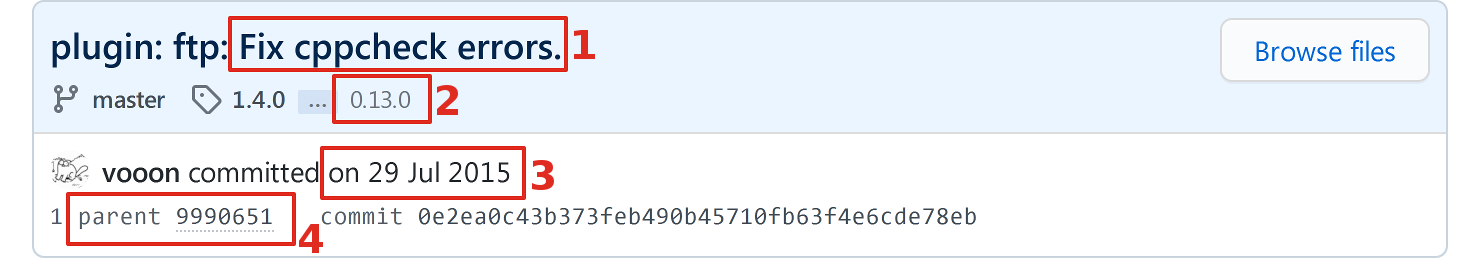}

  \end{center}

  \vspace{-2mm}

  \caption{Commit \texttt{0e2ea0c4} of \url{https://github.com/mavlink/mavros}.}%
  \label{fig:github-commit}

\end{figure}% __]

Inspecting commits and pull requests that do not reference any issue requires additional work, as they tend to describe \emph{what} changes are introduced, rather than the problem addressed.
Despite this, from the commit in \cref{fig:github-commit}, we can sill harvest four relevant items: The \emph{commit message}\,(1) includes the keyword \emph{``fix''}, implying there was an underlying issue. In this case, the commit fixes warnings from the \textsc{cppcheck} code analyzer.
The first \emph{release}\,(2) of the repository, which includes the commit.
Together with the \emph{commit date}\,(3) this determines the versions of the involved software.
The \emph{parent commit}\,(4) is the last version of the code that still contains the bug fixed.
\looseness -1

\paragraph {Data Analysis.}

For each of the bugs in the data set we produced a \emph{forensic description} by manually analyzing the available information.
Each description follows a common schema.
The initial list of attributes in the schema has been identified in a discussion of the authors based on their expertise in bug studies and in robotics software engineering.
The list has remained stable for most of the data collection period, but several fields have been added in an exploratory fashion.
These were usually derived either from the initial fields (using thematic coding) or by automatically querying GitHub repositories. Each description has been initially written by a team member familiar with the associated subject system, before being discussed extensively and cross-checked by multiple members of the research team.
We include all these descriptions as YAML documents, along with the schema in \robust\ repository on GitHub.
\looseness -1

\begin {figure} [t!]%[__

\begin{lstlisting}[
    firstline = 3,
    language = yaml
  ]
id: e964bbb
title: Calculation Error Inverts Turning Direction
description: "Due to an error in velocity calculations, when Kobuki was issued a 
  very low negative linear speed (slow backwards movement), it would also 
  inadvertently invert its turning direction. That is, if it was supposed to move 
  backwards while turning left, it would move backwards and turn right instead."
^\textbf{CWE}^: 
  id: 242
  description: Incorrect Calculation
keywords: ['differential drive', 'velocity', 'driver', 'movement']
system: kobuki
failure-codes: [ 'PROGRAMMING:CALCULATIONS' ]
fault-codes: [ 'SYSTEM:UNINTENDED-BEHAVIOUR', 'SYSTEM:MOTION' ]
bug:
  phase: runtime
  architectural-location: application-specific code
  task: locomotion
  package: yujinrobot/kobuki/kobuki_driver
  detected-by: developer
  reported-by: contributor
  issue: https://github.com/yujinrobot/kobuki/issues/227
  time-reported: 2013-02-22T09:09:21Z
fix:
  ^\textbf{commits}^:
    - repo: https://github.com/yujinrobot/kobuki
      hash: e964bbb8700fb1a9b95c0cfe5a44d43321294d4f
  pull-request: null
  fix-in: ['kobuki_driver/src/driver/diff_drive.cpp']
  languages: ['C++']
  time: 2013-02-25 (09:31)
\end{lstlisting}

  \caption {An example report for a bug (\bugkobuki {e964bbb})}%
  \label {fig:bug-report}

\end {figure}
% __]

\Cref {fig:bug-report} shows an example of a description for a bug in the \kobuki\ project.
It opens with a unique identifier, a prefix of the hash of its fixing commit in a Git repository (e.g., \texttt{e964bbb}).
The \texttt {title} summarizes the bug in general terms, and the \texttt {description} elaborates on the bug itself, the software components affected, and the context in which the bug occurred.
We wrote the descriptions aiming to be as accessible as possible, without presupposing deep training in robotics.
The \texttt {keywords} aid the search and retrieval of relevant bug reports.
(Unlike the codes discussed below, the keywords are not derived systematically.) The \lstinline {system} field records the name of the project in which the bug has been found.
\looseness -1

We initially attempted to classify bugs using \emph{Common Weakness Enumeration}, an established taxonomy of software weaknesses independent of us.\footnote{\url{https://cwe.mitre.org}}
However, as CWE is predominantly concerned with security, we were unable to adequately classify most of the dataset.
Motivated by this inadequacy, rather than re-using an existing taxonomy (e.g., IEEE 1044--2009; \citealt{Seaman08,Thung12,Garcia20,Wang21,Zampetti22}), we elected to use open coding and grounded theory building as an established mechanism for structuring qualitative data, when no prior taxonomy is pre-supposed. This allows us to better represent and fully describe the nature of software bugs in ROS without being constrained by an existing categorization. Moreover, by allowing the taxonomy to emerge from the data, our study provides a conceptual replication of prior work.

We systematically analyzed all bugs through a process of thematic coding by establishing codes in two groups: \emph{failure} descriptions and \emph{fault} descriptions. {\color{orange}} We define failure as inability of software to perform its function, with a special focus on the observable manifestation of this inability (sometimes also referred to as \emph{error}). The \emph{fault} is the cause, or the reason for the failure, within the software (so if the fault is repaired, the failure is eliminated). The results of this analysis are stored under \texttt {failure-codes} and \texttt {fault-codes} respectively.

The thematic coding has been split among the five coauthors randomly; two coders per each bug description. They performed the initial coding independently, introducing new codes as necessary.  After the initial coding has been obtained, we held a consistency meeting which produced a unified codebook.  Afterwards all bug descriptions have been recoded according to the codebook.  Finally, for all code assignments where the two coders disagreed we held a series of consensus meeting with all five coders---an agreement was achieved by a joint discussion, analysis of the source material, and any necessary context information about ROS.\@ Two of the coders involved had extensive robotics engineering experience, and three had extensive software quality engineering experience.
\looseness -1

The rest of the record is broadly split into two sections: the \emph {bug description} that elaborates on the fault and failure, and the \emph {fix description} that collects  information on how  the bug has been fixed.
The bug description specifies: the \emph{stage} at which failure occurs (e.g., build, deployment, runtime);
the relationship of the person that \emph{reported} the bug to the affected system (e.g., guest user, contributor, maintainer, automatic, unreported);
the URL of the associated GitHub \emph {issue}, and at what time the issue was reported;
the \emph{task} of the robot that is directly affected by the bug (e.g., perception, localization, planning);
a determination of how the bug was \emph{detected} (e.g., build system, static analysis, assertions, runtime detection, test failure, developer);
and whether the failure occurred in the application or in the ROS/ROSIn platform itself (\emph{architectural-location}). The fix description provides: a list of commits that constitute the bug fix; the URL of the associated pull request, the date and time at which the bug was fixed; the files that were changed as part of the fix, and the language of those files. We only include the subset of files that were changed and which relate to the bug fix itself. We do not include coincidental changes (e.g., refactorings).
\looseness -1

In \Cref{sec:analysis}, we give clickable links to bugs in the repository.
These lists of links are not exhaustive, in the sense that they show a small number of examples, not all the examples from the dataset in the given category.

\paragraph {Descriptive Statistics of the Obtained Dataset.}

Table\,\ref{tab:subjects} lists how many bugs we collected for each of the subject systems, and out of how many issues they have been selected (the remaining issues did not report bugs, so the statistics paint a valid picture of the bug population for this systems at the collection time).\looseness -1

\begin{figure}[t]% [__

  \setlength \HEIGHT {45mm}
  \begin {minipage} {.45 \textwidth}

    \vbox to \HEIGHT {
      \includegraphics [
         scale = 0.40
      ] {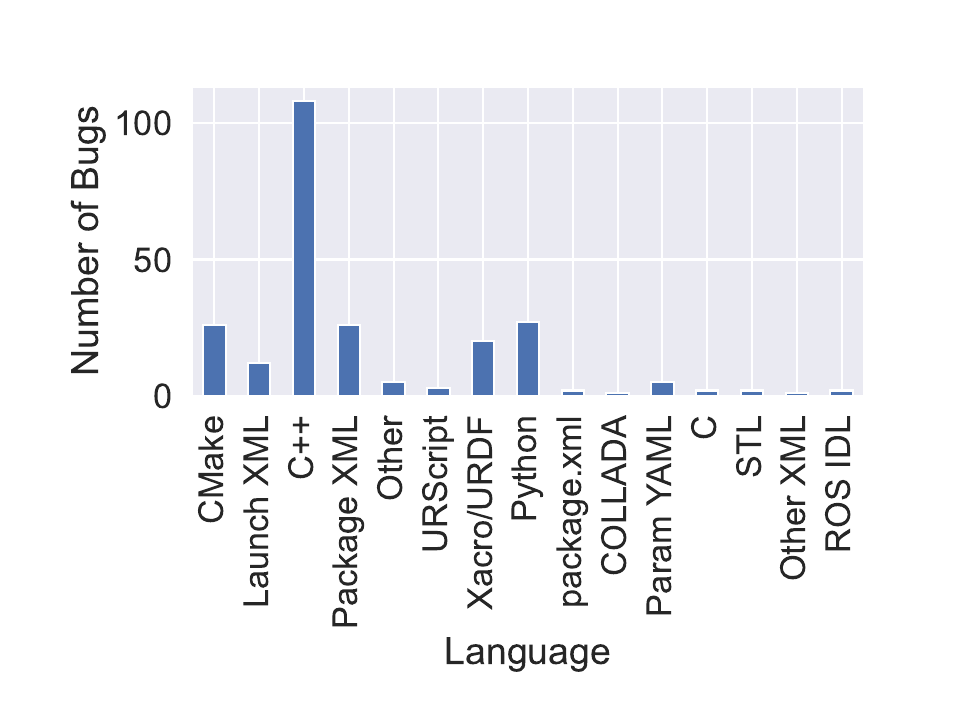}
    }

    \caption{The languages and file formats involved in bug fixes}
    \label{fig:language-vs-number-of-bugs}

  \end {minipage}
  \hfill
  \begin {minipage} {.5 \textwidth}

    \vbox to \HEIGHT {
      \includegraphics [
        scale = 0.40
      ] {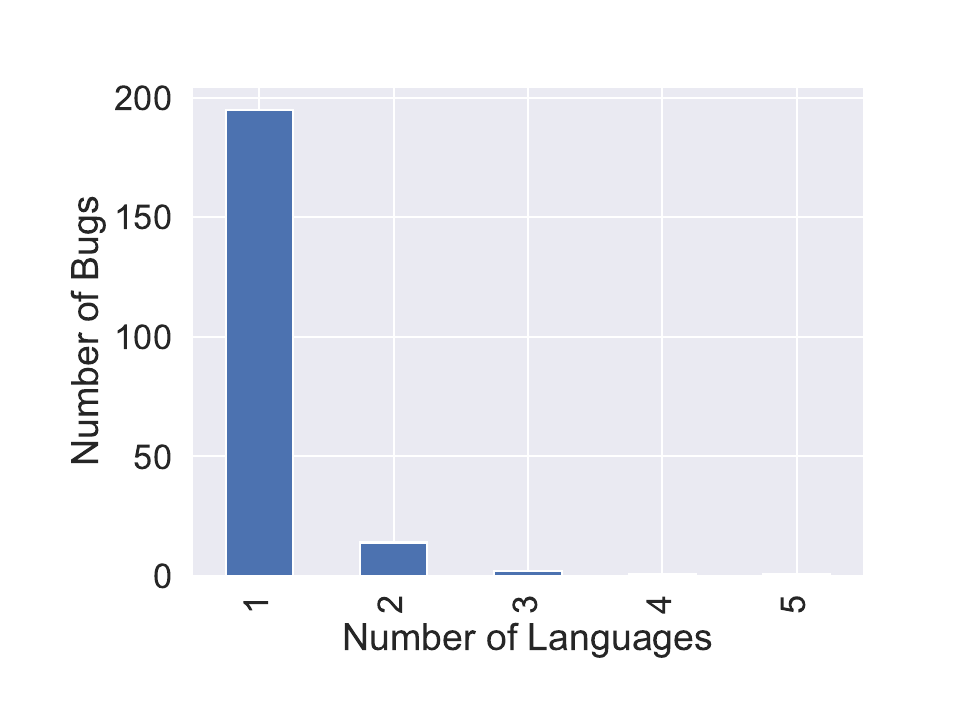}
    }

   \caption{The number of languages involved in bug fixes}
   \label{fig:number-of-languages-vs-number-of-bugs}

  \end {minipage}

\end{figure}

\Cref{fig:language-vs-number-of-bugs} breaks down the bugs by languages used in the fixed files, for the bugs that have been fixed.\@
Over half of the bug fixes involve C++ (112 of 219).
The remaining 107 fixes use a diversity of languages, many of which are domain-specific (e.g., Package XML, Launch XML, URScript, etc.), which typically lack associated analysis tools. \Cref{fig:number-of-languages-vs-number-of-bugs} shows the number of languages involved in each fix.
We find that 200 fixes (91\%) are limited to a single language.
That is, while failures may span components written in different languages, fixes are usually restricted to a single language.

In the dataset, \numruntimebugs failures occur at run-time and \numstartuptimebugs at start-up time, so in total \numruntimeandstartuptimebugs bugs that have been fixed also have execution-time failures.
Only \numfixedbugswithtests of these are accompanied by a test case.
(We automatically identified bug fixes that add or modify tests by checking the paths of the changed files.  We consider that a fix commit is accompanied by a test case if it adds or changes a file that contains the word \texttt{test} in its path, e.g., \texttt{test/foo.py}, \texttt{test\_foo.py}. We manually inspected the remaining fixes to confirm that they did not include a test case.)
\looseness -1

% __]

\begin{figure}[t]% [__

  \setlength \HEIGHT {50mm}
  \begin {minipage}{.5 \textwidth}

    \vbox to \HEIGHT {
      \includegraphics [
        scale = .40
      ] {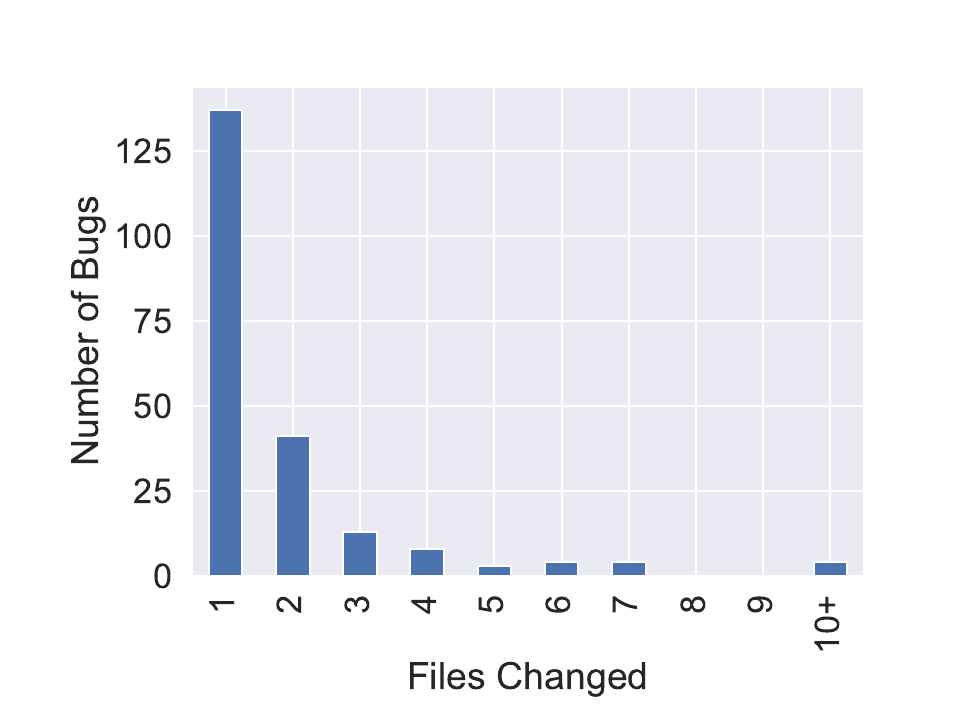}
    }

   \vbox to 5 ex {
     \caption {The number of files involved in a bug fix}
     \label {fig:number-of-files-vs-number-of-bugs}
   }

  \end {minipage}
  \hfill
  \begin {minipage} {.45 \textwidth}

    \vbox to \HEIGHT {
      \kern 2.7mm
      \includegraphics [
        scale = .39
      ] {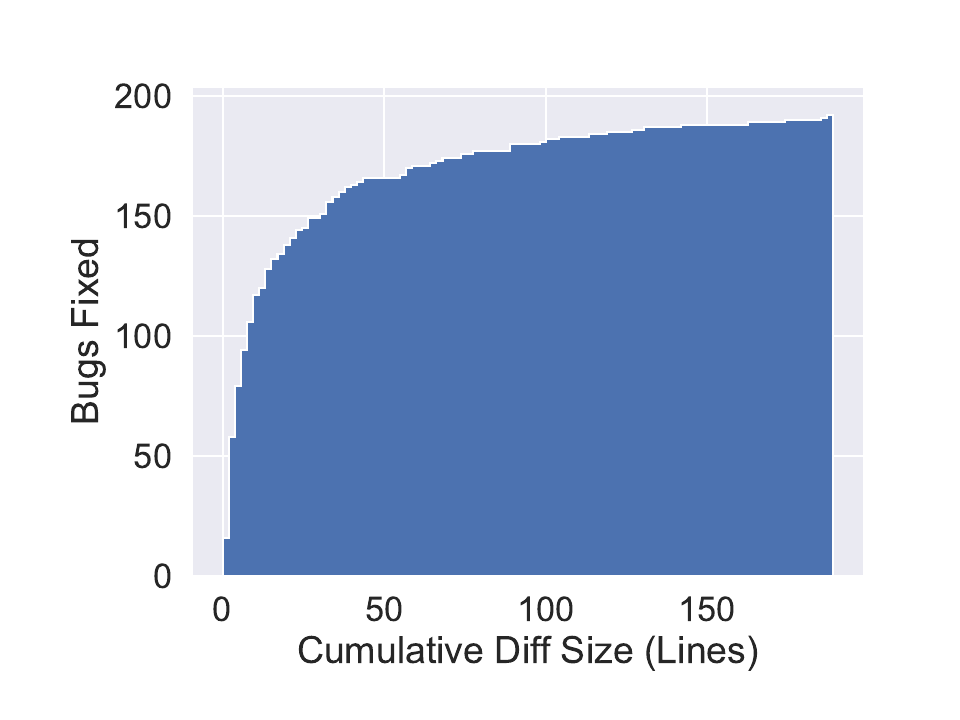}
    }

   \vbox to 5 ex {
     \caption {The diff size of a bugfix; we truncated 11 bugs with size above 200 lines\looseness = -1}
     \label {fig:change-difference-size}
   }

  \end {minipage}

  \bigskip

\end{figure}% __]

\Cref {fig:number-of-files-vs-number-of-bugs} shows an overview of the number of files that were fixed for each bug. The number is based on a manual removal of unrelated changes from bug-fixing commits. Almost two thirds of bug fixes affect only a single file. Specifically, 64\% of bug fixes are confined to a single file (141 of 219), 19\% span two files (41 of 219), and 17\% (37 of 219) change three or more files.

In order to approximate the size of each bug fix, we measure the number of lines in the change differences across their fixing commits; see \cref{fig:change-difference-size}. As fixing commits may contain unrelated changes (e.g., opportunistic refactoring), the size of the change difference is greater than or equal to the size of the bug fix, and therefore represents a conservative upper bound. Four bug fixes consist solely of file renamings and have an associated diff size of zero.  We find that more than 50\% of bug fixes have diffs that consist of 12 lines or fewer, and 75\% have a change difference that is 50 lines or smaller.

\paragraph{Limitations.}%
\label{sec:limitations}

The \emph {Credibility} \citep{Shenton04QualTrust,Sikolia2013TrustworthinessOG} of this study has been ensured by careful selection of the systems to be analyzed (\cref{sec:subjects}), by depending on qualified researchers for bug selection and analysis, by the use of established methods for both archival and coding of bugs (\cref{sec:method}), by employing peer scrutiny, and by grounding the study in existing work (\cref{sec:related}). All authors were involved in bug gathering, selection, and analysis.
One author is a domain expert on ROS with considerable experience in this FOSS community, the others have prior experience with ROS and with the kind of studies presented in this paper.
All bugs were analyzed by at least two authors and the results were cross-referenced and discussed among all.
Corrections were made by consensus.
The coding was done according to established practices~\citep{Saldana2015,linneberg2019coding} and by all authors in parallel, in multiple sessions.
The final classification of bugs and the resulting codebook have been extensively discussed and checked for consistency by all authors. Finally, the preliminary results of the study were presented at ROSCon, the main conference of the ROS community \citep{ROSConTimperley2019}.  Received comments were taken into account while further refining the study.
\looseness = -1

We used purposive sampling, so some negative impact on the \emph {Transferability} is expected.
The subject repositories were selected based on the role the packages have in ROS-based products.
Care was taken to select a qualitatively diverse set.
We do not claim that this set of packages is representative of the entire ROS landscape, or of the wider robotics software.
The quantitative results are not directly generalizable to these wider contexts---they have a been presented as descriptive statistics of the dataset, not as general conclusions.
Even with these restrictions however, the set of bugs described in \robust\ is diverse enough to be representative of the types of systems that were analyzed.  Conclusions can be made qualitatively about the presence of particular \emph{kinds of bugs} in other ROS packages.  Furthermore, while the selection of repositories was purposeful, the identification of bugs and fixes was not: bugs and fixes were always reported and contributed by either developers, maintainers or users of the systems under analysis, not by the authors.  The developers were treated as historical oracles and their assessments whether something is a bug, a fix or neither were taken at face value.
\looseness = -1

To increase \emph {Dependability}, we detail how we gathered the dataset (\cref{sec:subjects,sec:method}), how the historical images were build (\cref {sec:infrastructure}), and how bug reports are structured, analyzed (\cref {sec:method}) and stored (\cref {sec:infrastructure}).  All bug reports link back to the source data for each bug: affected source code repositories, the original issues, code contributions fixing the bug, and the state of the involved repositories \emph {before} and \emph {after} merging the fix.  Bug reports also include timestamps for these events, whenever they were present and identifiable, and the full analysis as performed by the authors.  All of the source material is made available open-source, on-line.  Such traceability increases dependability, facilitating evaluation of the research methods and results \citep {Shenton04QualTrust}.
\looseness -1

To warrant \emph {Confirmability}, the dataset was built from issue reports written by developers, maintainers, and users.  Some of these reports are unambiguous and leave no margin for researcher bias or interpretation, mostly because of the detail of the report and the language used by the reporter.  For instance, bugs related to the build phase of the software can hardly be mistaken for runtime issues.  Others contain little description of the fault or the manifested failures.
To minimize bias, we involved all authors in the analysis of such cases, relying on the buggy source code and the fix, until a consensus was reached, often over several meetings.
The commit history of the \robust\ repository, can be used to reconstruct how our bug reports changed over time, as a result of repeated analysis and discussion.
Other aspects of our method can also be audited, as it is fully explained and all source code is available online.
Despite this, we admit that a minimal set of bugs could be classified differently by another party for two reasons.
First, we did not interview issue reporters to confirm whether our judgement matches theirs.
Second, relying on the source code, especially on the changes introduced by a bug fix, might not tell a clear story because some commits might affect code that is not pertinent to fixing a particular bug.
After completing our analysis, we estimate the accuracy of our labels by sampling 30 bugs from the dataset and recoding them according to our final taxonomy.
Comparing the differences between the labels, we find that the original taxonomy had missing labels for three bugs and incorrect labels for a further three bugs, yielding a bug-level accuracy of 80\% across our sample.
Note that, only one label was incorrect or missing for each of those bugs (out of an average of 3.47 labels), giving us a label-level accuracy of 97\%.
\looseness -1

\section{Analysis}%
\label {sec:analysis}

\looseness -1
We analyze the collected data to understand what bugs developers experience in practice in robotics systems.
\@ The discussion is organized along two questions:
\begin {description}
\item [\textbf{RQ1:}] What software \emph {faults} occur in robotics systems?
\item [\textbf{RQ2:}] What \emph {failures} occur in robotics systems?
\end{description}

\noindent
We ask RQ1 to determine the extent to which the faults in the data are specific to ROS and robotics, identify common ROS development pitfalls, and provide valuable insights to guide the construction of effective QA tools. We ask RQ2 to gauge the actual and potential consequences of software failures in open-source robotics software, and to understand what QA tools are needed to automatically detect such failures. By assessing the extent of potential failures we hope to motivate further constructive research on robotics software quality. The analysis follows the open thematic coding method described in \cref{sec:method}. It gives a good qualitative description of the contents of the \robust\ dataset.

\begin {table}[b!]% [__ tab:themes:faults

  \caption{Top-level themes identified in fault analysis of \robust\ bugs.  Column\,\textbf{N} shows the number of bugs labeled with a given theme. Bugs may be labeled with more than one theme.}%
  \label{tab:themes:faults}

  \renewcommand \arraystretch {1.45}
  \renewcommand \tabcolsep {1 mm}

  \vspace {-2mm}

  \begin{tabularx}{\linewidth}{%
      @{}
      >{\footnotesize\bfseries\raggedright}p{20mm}
      >{\small}X
      >{\small}c@{}
  }

  Theme
  & \textbf{Description}
  & \textbf{N}
  \\
  \toprule

  Build, deployment, orchestration
  &
  Faults in source code and build infrastructure that ultimately lead to failures at build, deployment and composition-time.  This theme covers syntax errors, bad imports, broken dependencies, etc.
  &
  \numbdo
  \\

  Run-time configuration
  &
  Faults that occur in configuration files that lead to a misconfiguration of the system. Includes incorrect parameters, arguments, constants, topics, and namespaces.
  &
  28
  \\

  Concurrency
  &
  Faults related to the use of shared resources, including missing or flawed synchronisation, mistimings, and incorrect signal handling.
  &
  20
  \\

  Evolution
  &
  Faults that are introduced by the change of a component, whether that be internal or external, that is not handled by the rest of the code. This include changes in programming language, directory structure, internal and external APIs, and underlying hardware and firmware.
  &
  37
  \\

  General \hbox{programming}
  &
  Faults that typically occur during programming that are not exclusive to ROS or robotics. This includes incorrect logic and calculations, syntax errors, broken contracts (e.g., API and protocol misuses), and missing features.
  &
  123
  \\

  Models
  &
  Faults affecting the accuracy and consistency of the robot's model of the world.  This includes inaccurate robot and world models, missing or faulty transformations (e.g., coordinate frame transformations) and conversions (e.g., degrees to radians), and incorrect physical units.\looseness -1
  &
  20
  \\

  Systems
  &
  Faults that either occur outside of the source code and configuration files of a system, or due to the interaction between the software and the system that it runs on.  These faults may come from hardware devices and firmware, the system environment, or platform incompatibilities.
  &
  8
  \\

  \bottomrule
  \end{tabularx}

\end{table}% __]

\subsection{RQ1: What software faults occur in robotics systems?}%
\label{sec:analysis:rq1}

\Cref {tab:themes:faults} outlines the top-level themes that emerged from the thematic analysis of the software faults in our dataset. We discuss them in detail below.
\looseness -1

\paragraph{Build, Deployment, and Orchestration.}

ROS software is structured in \emph{packages} distributed with a package manager (e.g., \texttt {apt}, \texttt {pacman}, \texttt {dnf}) and in a source form. Packages are built using one of ROS's generic build tools, such as \texttt{catkin}. Under the hood, packages with \CC\ code are built with \texttt{CMake} using an accompanying \texttt{CMakeLists.txt} script obeying several ROS-specific conventions. Faults can appear in all specifications and scripts in the set up of packages, installation manifests, interface description languages (IDLs), and build. We include static compilation and linking errors under this theme as part of build~\examples{\bugmavros{282c9be}, \buguniversalrobot{39eb24f}, \bugmotoman{ddc6f36}}. In total, the \robust\ dataset contains \numbdo entries of this kind.\@ Interestingly, this is the second largest theme within the dataset, suggesting that these robotics systems suffer from modularity, dependency, and distribution issues significantly. In fact, these issues are more prevalent in these systems than the robotics-specific issues. Below, we discuss the sub-categories within this theme.
\looseness -1

\begin{itemize}
  \item \emph{Problems with code generation:} ROS automatically generates language bindings for clients using custom messages, services, and actions, from so called \texttt{.msg}, \texttt{.srv}, and \texttt{.action} files.  The generators or the generated code need to be available for the client and server code to build. Typically the errors relate to missing dependency of the build on the generator or to missing imports of the generated files \examples{\bugkobuki{17560e9}, \bugcareobot{105dc16}}.

  \item \emph {Dependency problems:} A \emph {manifest} \texttt {package.xml}\footnote {\url {http://wiki.ros.org/catkin/package.xml}} specifies dependencies for build-, test-, run-time, and documentation of a ROS package (\cref {fig:package-xml}).
  A dependency is either another ROS package (e.g., \texttt {std\_msgs}, \texttt {roscpp}) or a \emph {system dependency} installed  via a package manager (\texttt {apt} on Ubuntu).
  The dataset contains numerous dependency faults both for package \examples {\bugmotoman {9df36cb}, \buggeometry {e12e723}} and system dependencies \examples {\bugcareobot {c8091b6}, \bugcareobot {ac6a181}} that cause failures at build and execution time.  A \emph {missing run-time dependency} on \texttt {eigen\_conversions} in the \texttt {depth\_image\_proc} package   crashes the TurtleBot's image processing node, causing no images to be received from the camera \examples {\bugturtlebot {f01d952}}.  \emph {Transitive} dependency faults are when a package wrongly relies on its dependency to provide another one.
  Kobuki developers expected that \texttt {libusb-dev} includes a required header file \texttt {stdint.h}, which is only packaged with \texttt {libusb-dev} on some systems \examples {\bugkobuki {0027b57}}.  \emph {Circular dependencies}, where two packages depend on one another, can also cause inconsistent build outcomes and failures \examples {\bugkobuki{f95c384}, \bugturtlebot {3390789}}.
  \looseness -1

\item \emph{Problems with meta-packages:} Meta-packages simplify simultaneous installation of multiple related packages.  They solely contain dependency specifications.  ROS packaging standards do not allow proper packages to depend on meta-packages.  The build tool, \texttt{catkin}, enforces this rule by reporting an error\footnote{\url{https://ros.org/reps/rep-0127.html}} \examples{\buguniversalrobot{3a48064}, \buguniversalrobot{a58f4b5}}.

\item \emph{Problems at the ecosystem level:} In rare cases, dependency problems appear outside the artifacts of the individual packages.  In the case of \bugkobuki{e3c9bbc}, the fault was placed in the index of the ROS Distribution.  A \texttt{source} version of a package was missing from the index, preventing others from downloading and building the package from source.

\end{itemize}

\begin {figure} [t]% [__ fig:package-xml

  \begin{lstlisting}[language=XML]
<package format="2">
  <name>foo_core</name>
  <version>1.2.4</version>
  <description>This package provides foo capability.</description>
  <maintainer email="ivana@willowgarage.com">Ivana Bildbotz</maintainer>
  <license>BSD</license>
  <url>http://ros.org/wiki/foo_core</url>
  <author>Ivana Bildbotz</author>
  <buildtool_depend>catkin</buildtool_depend>
  <depend>roscpp</depend>
  <depend>std_msgs</depend>
  <build_depend>message_generation</build_depend>
  <exec_depend>message_runtime</exec_depend>
  <exec_depend>rospy</exec_depend>
  <test_depend>python-mock</test_depend>
  <doc_depend>doxygen</doc_depend>
</package>
  \end{lstlisting}

  \caption{A Package XML file  provides basic package metadata, including name, version, and dependencies (source: \href{http://wiki.ros.org/catkin/package.xml}{\textsf{wiki.ros.org/catkin/package.xml}}, 2020-08-10)}%
  \label{fig:package-xml}

\end{figure}% __]

\paragraph{Run-Time Configuration.}

ROS supports developers to build complex robot systems by composing reusable software components (nodes).
To reduce spatial coupling\,\citep{many-faces-pubsub}, and to facilitate rapid prototyping, the nodes API is unaware of the nodes responsible for providing, reading from, or writing to a particular topic. Instead, nodes interact via topics and their associated names. The use of strings to identify communication channels allows nodes to be spatially decoupled, thereby facilitating rapid prototyping and dynamic architectural changes. However, such \enquote{stringly types} \citep{stringlytypes} prevent, for the most part, misnamings and misconfigurations from being discovered until run-time \examples{\bugmavros{263650d}}.
The dynamic, stringly-typed nature of ROS's run-time architecture forces nodes to rely heavily on conventions and assumptions, which are typically neither enforced nor checked.
\looseness -1

To allow nodes to be used in a variety of contexts, configuration files often use name remapping and namespaces. Remappings (e.g., of topic names) are used to change names for a particular node, allowing nodes with different naming assumptions to interoperate. Unfortunately, incorrect remappings are easily introduced and sometimes difficult to identify \examples {\bugturtlebot {3e32933}}.
Topic remappings are often used together with namespaces to provide access to a group of related resources.
Namespaces are typically used to safely manage multiple instances of a particular component (e.g., a robot, sensor, algorithm) by creating a scope.
The use of hard-coded global names prevents multiple instances for a component and is considered an anti-pattern\,\examples {\bugturtlebot {a482f82}}.
\looseness -1

Parameters are also used to tailor the behavior of a component for a particular application or purpose.
Failure can occur when incorrect values are supplied to parameters \examples {\bugkobuki {d9aa656}} or when the wrong parameter name is used~\examples {\bugkobuki {8a729db}}.
If a node attempts to read from an undefined parameter (e.g., due to a typographical error), ROS may quietly use a default value, leading to unexpected and difficult-to-debug behavior \examples {\bugmavros {e1a8005}}.

\paragraph{Concurrency.}

ROS nodes can be implemented as either individual processes on the same or
different machines that intercommunicate via network protocols
(e.g., XMLRPC, TCPROS, UDPROS), or as threads within a single process
that intercommunicate via zero-copy messaging.
Naturally, ROS's distributed architecture leads to various concurrency and timing-related
issues.

The most common concurrency-related fault within our dataset is a lack of synchronization \examples {\bugkobuki {62a38a9}, \buggeometry {15b2e3c}, \bugmavros {1f01916}}. In some cases, synchronization primitives are present but are either misused, leading to liveness failures
\examples {\buggeometry {74f0c66}}, or are incomplete, failing to provide synchronization generally \examples {\bugroscomm {ca23e58}}.
We also observe a small number of timing-related issues within the dataset \examples {\bugkobuki {f548cc7}, \bugkobuki {5a44ead}}.
For example, when reporting the states of the robot's joints to the \texttt {joint\_states} topic, \bugmavros {753226d} failed to assign a timestamp to the \texttt {sensor\_msgs/JointState} message, leading recipients to ignore the message as stale.
Finally, faults may involve missing or incorrectly implemented signal handlers \examples {\bugmavros {29af3a3}}.
For example, in \bugkobuki {f548cc7}, a lack of appropriate signal handlers prevented the robot from safely and gracefully terminating its software processes upon receiving a \texttt {SIGTERM} signal.
\looseness -1

\paragraph{Evolution.}

Failures may suddenly appear as a result of changes to the environment in which the package is built and deployed, without any modification to its source code.
Similarly, internal changes in one part of a package may not be reflected in other areas
of that package leading to a variety of build issues and run-time failures
\examples{%
\buggeometry{0481047},
\bugmavros{bdda1fa},
\bugkobuki{8a729db}}.
Potentially disruptive changes may occur through the introduction of
newer programming language versions and compilers
\examples{%
\buggeometry{7677ca7}},
operating system distributions \examples{\bugturtlebot{928306b}},
and ROS distributions
\examples{%
\buguniversalrobot{56cf07f}},
which lead
to downstream issues in packages that rely on the older behavior.
Changes to the robot's underlying hardware and its associated firmware that are not reflected in its software may also lead to issues \examples {\bugkobuki {b18f559}, \bugmavros {de2cc36}}.
\looseness -1

Most commonly, problems arise as a result of changes to a dependency. While ROS packages may state their dependencies, there is no first-class mechanism for pinning those dependencies down to a particular version or set of versions, cf.\ PyPI\,\citep{python-version-strings}.
A library may alter,
remove
or deprecate
parts of its API or ABI,
leading to build failures
\examples{\bugkobuki{5abe7d4}, \bugkobuki{9c8abeb}}
or unexpected behavior at runtime
\examples{\bugturtlebot{61a75df}, \bugkobuki{55e84a6}}.
Issues are especially likely to occur when such changes are not reflected
in the documentation \examples{\bugkobuki{9682b9a}}.
Existing source code, configuration, and data files may be changed
\examples{\bugcareobot{b826eae}},
moved
to a different location \examples{\bugmotoman{6a7a506}, \bugmavros{101c09b}, \bugkobuki{5abe7d4}},
or disappear unexpectedly from a dependency that provides them \examples{\bugkobuki{8c30446}}.
ROS names (e.g., topics, services, action servers, parameters) and namespaces
may be inconsistently changed between source code and configuration files
\examples{\buguniversalrobot{778c1ac}, \bugmavros{263650d}, \bugkobuki{8a729db}},
and particular
publishers, subscribers, and services may be changed or removed
\examples{\bugmavros{bdda1fa}}.

\paragraph{General Programming.}

Unsurprisingly, many faults within the ROS systems are general programming
mistakes that could occur in any software, including
typo mistakes, using the wrong logical or arithmetic
operator \examples{\bugmavros{c172409}},
\enquote{copy-paste} or \enquote{clone-and-own} mistakes \examples{\buggeometry{439e235}},
code smells \examples{\bugconfidential{86cb680},},
and issues stemming from \enquote{stringly types} \examples{\bugmavros{dab1b8a}}.
\looseness -1

We observe a variety of mathematical, logical, and control-flow issues.
Several bugs stem from incorrect loop invariants \examples{\bugmavros{86255ba}},
mishandled loop variables \examples{\bugmavros{215010d}},
and missing loop break conditions \examples{\bugkobuki{e34428d}}.
A number of bugs occur as a result of missing or incorrect input validation
\examples{\buggeometry{d12b890}}
and robustness measures \examples{\bugconfidential{2688e7a}},
fail to adequately account for certain corner cases and
boundary conditions \examples{\bugkobuki{af7946f}},
or lack important features
\examples{\bugmotoman{b1b6fcb}}.
Faults may also occur in mathematical calculations
\examples{
\buggeometry{860b866}
}.
Bugs also occur as a result of API misuse and contract violation
\examples{%
\buguniversalrobot{b3c2c21}},
or, conversely,
when an implementation does not satify its API specification
\examples{\bugmavros{599c588}}.

We observe a number of issues specific to \CC, for example:
string length
\examples{\bugmavros{2998e9f}},
string formatting \examples{\buggeometry{164cfa3}},
namespaces \examples{\buggeometry{b206807}},
multiple definitions \examples{\buggeometry{f19569c}},
and zero-copy messaging \examples{\bugkobuki{dbcdb12}}.
Uninitialized variables
\examples{%
\bugmavros{fcf9cd9}%,
}
and incorrect type casts and conversions
\examples{%
\bugmotoman{292b5cc}%,
}
can lead to surprising failures at run-time.
Numerous issues occur as a result of poor resource and memory management
including off-by-one errors
\examples{\buguniversalrobot{cda133d}, \buggeometry{729a653}},
overreading buffers \examples{\buguniversalrobot{359a2e9}},
and incorrectly calculated indices \examples{\bugkobuki{9397c6b}}.
Similarly, we also notice issues related to the misuse of Python features including missing or incorrect type conversions and checking \examples{\buggeometry{cec6208}, \buggeometry{e4466f0}, \buggeometry{d12b890}},
and bad imports \examples{\bugturtlebot{61a75df}}.
\looseness -1

\paragraph{Models.}

All of these faults lead to the robot forming an inaccurate model of reality.
They may occur as semantic errors in the robot's URDF, Xacro, STL,
and DAE files, which provide a physical and visual description of the robot for
motion planning, visualization, and simulation.
For example, the 3D meshes used to describe the robot may be missing, malformed,
or incorrectly handled
\examples{\bugmotoman{0829607}, \bugmotoman{6a7a506}, \bugkobuki{4ea5ea7}}.
Alternatively, the robot description files may
incorrectly specify physical dimensions, mass, and inertia of the robot
\examples{\bugkobuki{493e3f9}, \buguniversalrobot{21b86f6}, \bugmotoman{1ec8ca1}}.

\begin{figure}
  \centering
  \includegraphics[width=0.5\textwidth]{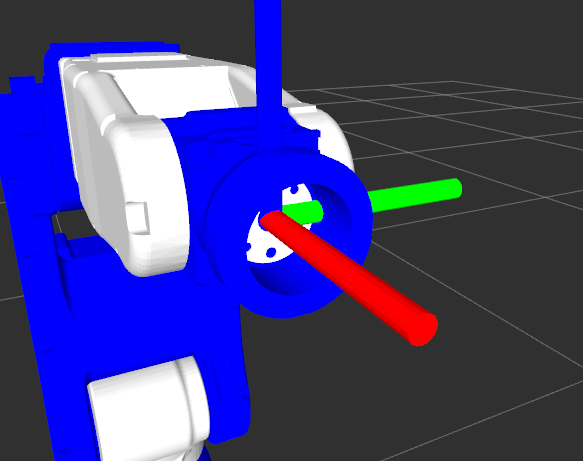}
  \caption{A subtle fault in the specification of a joint
  led to an inaccurate robot model being used for planning, visualization, and simulation \examples{\bugmotoman{1ec8ca1}}.}
\end{figure}

Inaccurate models of reality can also come about as a result of
missing or incorrect transformations and conversions \examples{\bugmavros{b96bf67}}.
For example,
\bugother{22e4e4f} sees the visual and depth data from a Kinect V2
camera and its associated Freenect2 driver
presented in a non-standard format,
flipped around its vertical axis (\enquote{mirrored}),
to an application that is unaware of the transformation.
Similarly, an error in a parameter name in \bugmavros{ff581a0}
sees the robot incorrectly report its (x, y, z) coordinates as
(x, x, x).
In bug \bugmavros{b96bf67}, MAVROS, which operates in a different coordinate frame to MAVLink, fails to correctly handle rotation when converting and sending coordinates to MAVLink.
\looseness -1

\paragraph{Systems.}

Faults occur as a result of an interaction between the software and the
system to which it is deployed. This includes the use of
operating system and distribution-specific code or file formatting,
preventing the software from being deployed to some platforms
\examples{\bugkobuki{95b24e8}, \bugmavros{31ad11d}, \bugturtlebot{a4f35ee}}.
\looseness -1

Alternatively, faults may occur due to interaction between the ROS software and
an underlying (faulty) hardware device and its associated firmware
(e.g., \bugkobuki{841720a}, \bugkobuki{b18f559}, \bugmavros{de2cc36}).
For example, in \bugkobuki{606b8b9}, when the USB serial cable is disconnected
from the robot, the ROS driver node will, following the \texttt{udev} rules for the robot,
attempt to read data from an unsupported bluetooth interface,
leading the node to crash under certain conditions.
\looseness = -1

ROS packages can also provide shell scripts known as environment hooks which are
typically used to set up environment variables. Issues may arise from the misuse
of environmment hooks. For example,
\bugturtlebot{3ea2c30}, introduced an environment hook that depended on
a package that was not stated as a dependency in the package manifest.
In cases where the other package was not coincidentally installed, error messages
would be printed to the terminal.

\subsection{RQ2: What failures occur in robotics systems?}
\label{sec:analysis:rq2}

We refine the research question RQ2 into the following sub-questions:
\begin{description}

  \item[\textbf{RQ2.1}] What \emph{immediate failures} occur in robotics systems?

  \item[\textbf{RQ2.2}] What \emph{ultimate failures} occur in robotics systems?

\end{description}

\noindent
Both questions address failures, but do so at different levels.
Immediate failures, or the \emph{software-level failures}, are failures that are immediately noticeable when testing or using software;
for example, a ROS node crashes at startup or sends messages at a slower rate than expected.
The propagation of the immediate failure throughout the entire robotic system might however result in a different kind of externally observable failure---the ultimate failure manifestation, or a \emph{system-level} failure.
For instance, a node that publishes incorrect velocity might make the robot turn to the wrong direction or become unresponsive, if other components detect and reject the incorrect values. Software-level failures rarely have no observable effect at the system level.

\subsubsection*{RQ2.1: Software-Level Failures}

\begin{table}[b!]

  \caption{Top-level themes of software-level failures identified within the dataset. Column \textbf{N} shows the number of bugs labeled with a given theme. Bugs may be labeled with at most one theme; no label is applied to bugs that do not result in software-level failure.}%
  \label{tab:themes:software-failures}

  \renewcommand \arraystretch {1.45}
  \renewcommand \tabcolsep {1 mm}

  \begin{tabularx}{\linewidth}{
      @{}
      >{\footnotesize\bfseries\raggedright}p{20mm}
      >{\small}X
      >{\small}c@{}
  }

  Code
  & \textbf{Theme description}
  & \textbf{N}
  \\
  \toprule

  Build
  &
  Failures that prevent the successful completion of the build or installation process, such as compilation and linking errors.
  &
  {\numbugsswbuild}
  \\

  User Experience
  &
  Failures that exclusively affect the user experience and do not result in software crashes or undesired behavior, nor affect the functional or non-functional performance of the robot, unless one considers the operator providing inputs to the robot via a user interface.
  &
  {\numbugsswui}
  \\

  Performance
  &
  Failures that manifest in the degradation of non-functional performance of one or more software components, but do not necessarily result in any observable degradation of functional performance at the system level.
  &
  {\numbugsswperf}
  \\

  Crashing
  &
  Failures that lead to a software crash in an individual node at startup or runtime, such as failing to locate runtime dependencies, memory corruption or simple type errors in dynamically-typed programming languages.\looseness -1
  &
  {\numbugsswcrash}
  \\

  Liveness
  &
  Failures that cause an individual software component to become unresponsive due to, for example, a deadlock or an infinite loop.
  &
  {\numbugsswliveness}
  \\

  Network
  &
  {Failures that affect messaging behavior (between components) of individual software components, resulting in undesired functional behavior in either the sender, the receiver or both.}
  &
  {\numbugsswnetwork}
  \\
  Behavioral
  &
  Failures that result in undesired functional behavior in one or more software components, and which do not lead to software crashes or liveness issues. Behavioral failures may be caused by, for example, miscalculations, logical errors, and violating contracts, or under certain unpredictable circumstances, memory-related errors (e.g., stack and heap corruption).
  &
  {\numbugsswbehavioral}
  \\

  \bottomrule
  \end{tabularx}
\end{table}

\Cref {tab:themes:software-failures} summarizes the high-level themes that resulted from the thematic analysis of immediate software-level failures in \robust. For each theme, we list the relative frequency in the dataset. We provide further details on each theme below.
\looseness -1

\paragraph {Build.}

Bugs that manifest at build- or install-time are most often a direct consequence of faults under the \emph{Build, deployment and orchestration} theme. Faults from other categories can also manifest at build time, when detected by code scanning tools or compilers. As they occur prior the system being able to run, they cannot result in a system-level failure.
\looseness -1

Two main types of build bugs exist in this category: (i) bugs which cause immediate build failures, and (ii) bugs which cause failures when packages are consumed as dependencies by developers or users. Examples of the former include not linking against used libraries\,\examples {\bugkobuki {ddc6f36}}, missing dependencies on code generator targets\,\examples {\bugcareobot {105dc16}}, violating packaging policies\,\examples {\buguniversalrobot {0c34123}}, exporting incorrect package metadata\,\examples {\bugkobuki {45ee84a}} and incorrectly specifying install targets\,\examples {\buggeometry {a723ecb}}. Examples of the latter include missing declarations of \emph{transitive} dependencies\,\examples {\buggeometry {4c160d3}} and an incomplete build environment setup\,\examples {\bugkobuki{fd6b589}}. Examples of deployment related bugs include \bugkobuki{9de9690} and \bugmotoman{259b468}, which both cause files expected by users to be absent from the runtime environment. The failures manifesting in dependent projects are particularly hard to diagnose, as they are observed by users in a completely different context than the one containing the fault.
\looseness -1

\paragraph {User Experience.}

Bugs in this category influence the efficiency and efficacy of a robot operator affecting the way in which an operator either receives information from or provides commands to the robot.

Examples include misreporting the status of hardware subsystems, such as battery state\,\examples {\bugkobuki {841720a}, \bugkobuki {bb3c7ec}}, not showing information due to misconfiguration of the interface\,\examples {\bugkobuki {8a729db}}, incorrect positioning of 3D data in the UI due to  parameter lookup falling back to defaults\,\examples {\bugmavros {84264f0}}, or because of updates to visualization tools which are incompatible with used 3D models\,\examples {\bugturtlebot {9299530}}.
Other example bugs include faulty data transformation causing wrong information to be shown to users\,\examples {\bugmavros {a67d81d}} and presenting the same data twice under different names due to use of an outdated communication protocol\,\examples {\bugkobuki{b18f559}}.
\looseness -1

\paragraph {Performance.}

Failures observed as a reduction of performance of individual components, but which do not influence the perceived performance of a complete system. Examples include not using language or framework features for efficiently passing data \examples{\bugmavros {0e2ea0c}, \bugkobuki {dbcdb12}} and misconfiguration of framework features resulting in less efficient passing of data \examples{\bugkobuki {dd40270}, \bugkobuki {5ee65b0}}.
\looseness -1

\paragraph {Crashing.}

Failures observed as a crash of an individual node at either startup or runtime.
The faults causing these failures include missing runtime dependencies (ROS nodes\,\examples {\bugturtlebot {891cb68}}, Python modules\, \examples {\bugcareobot {c8091b6}}, shared libraries\,\examples {\bugturtlebot {f01d952}}), incorrect linking of libraries\, \examples {\bugkobuki {3e88010}}, use of non-existent files \examples {\bugkobuki {9682b9a}, \buggeometry {fc854e0}}, use of malformed files \examples {\bugmotoman {90a9464}, \bugcareobot {b826eae}}, incompatible firmware on external devices \examples {\bugkobuki {c04eae5}}, management of resources \examples {\buggeometry {001fca6}, \buggeometry {6c13c78}, \bugmavros {215010d}}, out-of-bounds access \examples {\buggeometry {729a653}, \bugmavros {4fb6e7e}}, incorrect or unsafe concurrency \examples {\bugroscomm {ca23e58}, \buggeometry {12605ab}} and hard-coding expectations about OS or runtime environment\,\examples {\bugturtlebot {a4f35ee}}.

\paragraph {Liveness.}

These failures render an individual software component unresponsive due to, for example, a deadlock or an infinite loop. Liveness failures are a form of denial-of-service. They disturb the continued operation of a component or prevent the safe termination of the component or its service. Deadlocks occur both internally to a component\,\examples {\bugmavros {1f01916}, \buggeometry {74f0c66}} and between an external system \examples {\buguniversalrobot {bd1fce5}}. Infinite loops can be caused by incorrect handling of termination conditions \examples {\bugkobuki {38dce2a}, \bugmavros {86255ba}} or absence of such conditions \examples {\bugkobuki {6e748c1}}. Other faults causing liveness failures include incorrect use of network functionality \examples {\bugroscomm {eab0d3c}}.
\looseness -1

\paragraph {Network.}

Network failures manifest by affecting messaging between individual components, resulting in undesired functional behavior in either the sender, the receiver, or both.
Examples include incorrect handling of communication failures and protocol mismatches\,\examples {\bugconfidential {2688e7a}, \bugconfidential {332f09f}, \bugconfidential {c5dc9de}}, failing to set message fields which may lead to subsequent misinterpretation\,\examples {\bugmavros {594978d}}, missing publisher queue sizes leading to inefficient and unintended blocking\,\examples {\bugkobuki {dd40270}}, and messages being lost due to an incorrect implementation of IPv6\,\examples {\bugroscomm {eab0d3c}}. \looseness -1

\paragraph {Behavioral.}

This theme covers run-time failures that result in undesired functional behavior other than software crashes or liveness issues. Behavioral failures are caused by, for example, miscalculations\,\examples {\bugkobuki {af7946f}}, incorrect implicit assumptions about input\,\examples {\bugconfidential {96e2c6c}}, logic errors\,\examples {\buggeometry {566092b}, \bugmotoman {292b5cc}, \bugroscomm {ca23e58}}, contract violations\,\examples {\bugmavros {599c588}, \bugmavros {de2cc36}}, misconfigured networking\,\examples {\bugkobuki {dd40270}, \bugroscomm {eab0d3c}}, incorrect processing of network data\,\examples {\bugmavros {594978d}}, or memory-related errors (e.g., stack and heap corruption). Some of these are robotics-specific, but most are usual specification violations known from regular software development.
\looseness -1

\subsubsection* {RQ2.2: System-Level Failures}

Many faults initially manifest in a component and then propagate to other components, possibly affecting the system as a whole, giving rise to \emph {system-level failures}.
We analyzed the repositories for drivers, controllers, and other components that are, ultimately, building blocks for concrete robotic applications. But, given that there is no well-defined concept of \emph{system} or \emph{application} in ROS, what looks like a fully functional system (e.g., a robot patrolling a known map) can either be the complete system, as envisioned by the users, or simply a complex component of a larger system (e.g., multiple robots working in concert).
Thus for bug reports devoid of information at the system level, such as what was the robot's mission and how it went wrong, we resort to to our experience and make an educated guess at how software-level failures manifest from a system perspective.
\looseness -1

\begin{table}[b!]% [__
  \caption{Top-level themes of system-level failures identified within the dataset. Column \textbf{N} shows the number of bugs labeled with a given theme. Bugs can be labeled with more than one theme; no label is applied to bugs that do not result in system-level failure.}%
  \label{tab:themes:system-failures}

  \renewcommand \arraystretch {1.45}
  \renewcommand \tabcolsep {1 mm}

  \vspace {-2mm}

  \begin{tabularx}{\linewidth}{
      @{}
      >{\footnotesize\bfseries\raggedright}p{20mm}
      >{\small}X
      >{\small}c@{}
  }

  Code
  &
  \textbf{Theme description}
  & \textbf{N}
  \\
  \toprule

  Loss of Functionality
  &
  Failures that make the functionality offered by a group of system components unavailable. User's perspective: \emph{\enquote{the camera stopped working!}}
  &
  {\numbugslossfun}
  \\

  Unresponsive
  &
  Failures that make the system become unresponsive to a number of operator commands. User's perspective: \emph{\enquote{it does not move!}}
  &
  {\numbugsunresponsive}
  \\

  Degraded Performance
  &
  Failures that cause the system to miss deadlines or perform one or more tasks with reduced timeliness. User's perspective: \emph{\enquote{it moves too slowly!}}
  &
  {\numbugssysperf}
  \\

  Behaving Incorrectly
  &
  Failures that make the robot perform unintended movements or actions, or perform the intended commands but with unexpected outcomes or side effects. User's perspective: \emph{\enquote{it turned left instead of turning right!}}
  &
  {\numbugssysbehaviour}
  \\

  Monitoring
  &
  Failures that result in inaccurate observation and diagnosis of the system's current state, via user interfaces of various types. User's perspective: \emph{\enquote{I cannot see the battery levels!}}
  &
  {\numbugsmonitor}
  \\

  \bottomrule
  \end{tabularx}
\end{table}% __]

For RQ2.2, we performed a thematic analysis of the externally observable symptoms of failure at the system level. \Cref {tab:themes:system-failures} lists the five high-level themes that resulted from this analysis.
As we can see in the table, system-level failures are different from software-level failures. First, the definitions are more open-ended, in the sense that many types of observable symptoms fit under a given theme. Second, while the themes are distinct, there are no clear boundaries between themes, or a logical sequence over them.
It is possible for a software failure to propagate and manifest in such a way that it fits multiple themes at the system level.
For example, consider a robot driver that receives velocity messages, updates its estimated pose, and converts the velocity message into actuator commands for the hardware.
If the conversion is miscalculated (e.g., wrong sign), the driver would report a correct pose estimation, as per the received messages, but would move in unintended ways (e.g., turning right instead of left).
While this is clearly a behavioral failure (the robot does not do what it is told), it could also be perceived as a communication failure (the robot reporting a pose estimate that diverges from its observed movement).
\looseness -1

Not all faults propagate all the way up.
It is possible for software-level failures to be \emph{harmless} at the system level, or to have no manifestation at all.
For an obvious example, a failure during the build process does not affect the system as a whole, because the system is not yet deployed at this stage.
The \robust\ dataset contains \numbugssysnone{} bugs that had no meaningful manifestation at the system level.
An example of this is \bugkobuki {8163705}, in which developers handled a raised exception within the code, but reported the handled exception in the error logging stream instead of the debug stream.
This misled users into believing that an actual error was occurring.
Despite the faulty component, the system as a whole worked correctly.
We now provide further details on each of the identified failure themes.
\looseness -1

\paragraph {Loss of Functionality.}

This theme covers failures that bring down parts of the system.
Partial loss of functionality prevents the use of certain non-critical features under particular conditions. This is typically something the system can recover from even though it might affect its overall performance. Substantial losses, on the other hand, are more likely to result in mission failure. For example, if a robot's safety controller is unable to respond to bumper events it would not hinder the robot's ability to perform its mission, if it never bumps into an obstacle, or if a redundant sensor is present. But losing the safety controller in its entirety, would likely be considered too dangerous to keep the robot in operation.
\looseness -1

The software-level failures that most likely escalate to a loss of functionality are crashing and liveness failures. If a component, or group of components, crashes or becomes otherwise unresponsive, the functionality it provides is lost to the system. Behavioral failures can also result in a loss of functionality. When a component behaves incorrectly due to logical mistakes, a functionality may be locked away due to the program never entering the appropriate execution path. User experience failures propagate in a similar fashion; a poor user interface might omit features by missing buttons or using unclear language. Network failures are more varied; loss of functionality, in this case, could be due to dropped messages in a misconfigured system, where the channel that enables the functionality becomes unusable. Performance failures are less likely to lead to loss of functionality, unless features are lost when their response times become too slow.
\looseness -1

In the \robust\ dataset, most software-level failures that escalate to a system-level loss of functionality are crashes (20 out of \numbugslossfun {}).
The remaining bugs are split between liveness\,(2), network\,(8) and behavioral failures\,(12).
There are examples of both partial and substantial functionality loss.
In \bugmavros {86255ba} an infinite loop prevents users from accessing and manipulating parameter values at runtime, while in \bugroscomm {eab0d3c} the robot is unable to communicate with a ground control station (both partial losses).
Crashes in the robot driver prevent interaction with the robot base\,\examples {\bugmotoman {377d7be}, \buguniversalrobot {359a2e9}}, and a bad topic remapping causes velocity commands to be ignored, preventing all movement\,\examples {\bugkobuki {35682ec}} (critical functionality losses). Software crashes in the vision pipeline, an essential part of localization and navigation, can also prevent autonomous movement of the robot\,\examples {\bugturtlebot {f01d952}, \bugturtlebot {3e32933}}.
\looseness -1

\paragraph {Unresponsive.}

These dangerous failures make the system unresponsive from the user perspective. The human operator cannot control the robot, despite it being turned on and working. The operator might be unable to make the robot perform its mission and to react to unintended and potentially dangerous behaviors.
\looseness -1

Any software-level runtime failure can escalate to unresponsiveness.
A crash of a node can prevent operator inputs from reaching the robot.
A node that behaves incorrectly, say a velocity multiplexer, can prevent user input from reaching the robot due to logical errors.
The \robust\ dataset contains examples of unresponsiveness that occur deterministically at startup\,\examples {\bugkobuki {fbe70c7}, \bugmavros {263650d}}, unexpectedly at runtime\,\examples {\buguniversalrobot {359a2e9}, \buguniversalrobot {bd1fce5}}, or when the operator attempts to safely stop the robot\,\examples {\bugkobuki {38dce2a}, \bugmavros {29af3a3}}.
In terms of the software-level failures, 20 out of \numbugsunresponsive{} records fall under one of two sources: \emph {crashing} and \emph {liveness}.
In the former, robots may become unresponsive due to missing runtime dependencies that crash core components\,\examples {\buguniversalrobot {58790ba}}. As for the latter, robots may become unresponsive due to, e.g., infinite loops\,\examples {\bugmavros {86255ba}}.
In another example, the program ignores the termination signal, preventing the user from safely terminating the robot with \texttt{SIGTERM} if the robot begins to behave dangerously\,\examples {\bugkobuki {054c753}}.
\looseness -1

\paragraph {Degraded Performance.}

Safe and effective robot operation often relies on the system behaving in a timely manner. Some computations are expected to complete periodically, with a given frequency; others are expected as a timely reaction to a given stimulus. A system fails due to \emph{lag} (reduced timeliness) if it does not meet expected deadlines. For many robotic tasks, such as detecting and avoiding obstacles or picking up parts from a conveyor belt, a delay means failure.
\looseness -1

Any software-level runtime failure may reduce system timeliness.
A crash of a node that is set to \texttt {respawn} (i.e., start again after sudden termination) is a failure from which the system can recover, but likely with noticeable delays.
Intermittent liveness failures can easily lead to reduced timeliness, too.
Still, performance-related software failures (e.g., CPU and GPU load) are likely the main culprit of degraded non-functional performance for the system as a whole.
\looseness -1

The \numbugssysperftext software failures in \robust\ that may affect the timeliness of the system are of types \emph {crashing}, \emph {liveness}, and \emph {performance}. For example, \buggeometry {1b5fa94} leaks memory at the component level, at a rate of about 15 megabytes per second. When running for long enough, systems tend to slow down, due to the lack of resources, until they eventually crash.

\paragraph {Behaving Incorrectly.}

A robot's observable behaviour is described not only in terms of digital effects as in other software systems, but also in terms of physical effects, such as moving from one place to another. When the software produces unintended outcomes, such as moving in the wrong direction or at a wrong velocity, we face a functional failure. This type of failure can easily pose physical danger to humans, living beings, the robot itself, and surrounding valuable objects.
\looseness -1

Any software-level failure could end up escalating to incorrect functional behavior.
Crashes or liveness failures in runtime monitoring components, such as safety controllers, could make the robot move into objects. Performance failures in components that issue movement commands can prevent the robot from acting as intended.
User experience failures can also lead to unintended movement.
If a user interface assigns the wrong label to a command button, the user would observe a robot movement that does not match their expectations.
Finally, behavioral failures (especially those stemming from miscalculations or logical errors) are among the most likely causes for unintended movement.
\looseness -1

Many functional failures in the \robust\ dataset originate from behavioral software failures (36 out of \numbugssysbehaviour{}).
Failures may be caused by incorrect or imprecise calculations\,\examples {\buggeometry {439e235}, \bugkobuki {af7946f}}, missing frame conversions\,\examples {\bugmavros {248cb38}}, and the use of absolute rather than relative time\,\examples {\buguniversalrobot {b3c2c21}}.
In \bugkobuki {af7946f}, a calculation error causes the robot to move slowly and turn in wrong direction when dealing with low negative linear velocities and rotation commands.
Unintended movement may also occur as a result of rounding errors and numerical imprecision\,\examples {\bugmotoman {9bf25ea}, \bugkobuki {1c141a5}}. For example, in \bugkobuki {1c141a5}, a loss of precision while converting a \texttt{float} to a \texttt{short} during a repeated calculation caused the robot to move in the opposite direction in special cases.
Missing or incorrectly implemented safety features can also lead to unintended motion, such as in \bugmotoman {b1b6fcb}, where the robot is able to immediately resume motion from a paused state.
In \bugkobuki {ad906f0}, a lack of acceleration smoothing causes the robot to abruptly perform wheelies when instructed to move from an idle position. Lastly, hard-coded speeds in \bugkobuki {0416c81} prevent heavier-than-anticipated robots from being able to move, and inaccurate joint limits in \buguniversalrobot {89145c4} and \bugmotoman {2d42582} cause the robot arm to collide with itself or attempt to reach physically unreachable positions.
\looseness -1

\paragraph {Monitoring.}

These failures hide or misrepresent the robot's internal state from a user perspective. They prevent the human operator from accurately observing or diagnosing the robot.
Being able to observe the state of the robot is crucial to ensure an additional layer of safety and emergency responses from the human operator (explainability).
Accurate monitoring may help prevent a robot from hitting an obstacle or overloading hardware components.
\looseness -1

Almost any failure can lead to divergence between the actual state of the robot and what the operator observes.
Incorrect calculations in a sensor node can manifest in incorrect behavior (publishing wrong data), which could cause the system to build and display an incorrect model of the world.
Node crashes, node liveness issues, or degraded non-functional performance likely result in stale data and a model that lags behind the actual state of the robot and its environment.
\looseness -1

Our dataset contains mostly examples of incorrect \emph{behaviour} that escalate to monitoring failures (9 out of \numbugsmonitor{}). For instance, in \buggeometry{860b866} incorrect vector translations end up producing wrong pose values, which ends up producing incorrect visualizations. In \bugmavros{fcf9cd9} and \bugmavros{753226d} issues with visualization are caused by uninitialized variables carrying wrong default values.
\looseness -1

\section{Findings}\label{sec:findings}

In this section, we interpret the results of our study in light of the existing literature and discuss the implications for both practitioners and researchers. \Cref{tbl:key-findings} summarizes the key findings and their impact on future work.

\begin {table}[!b]% [__

\caption {A summary of the key findings from building the \robust\ dataset}%
\label {tbl:key-findings}

\renewcommand \arraystretch {1.30}
\renewcommand \tabcolsep {1 mm}

\vspace {-2mm}

\begin {tabularx} {\linewidth} {%
    @{}
    >{\small}p{48 mm}
    >{\small}X@{}
}
    \textbf {Finding}
    &
    \textbf {Impact / Action / Future Work}
    \\

    Many late-stage failures are caused by misconfiguration, feature interactions, and co-evolution of components, platforms, and hardware.
    &
    There is need for better tools to detect and address potential incompatibilities between hardware and software components ahead of deployment.
    Our dataset provides many case studies for researchers willing to address this challenge.
    \\

    Many of the problems contained in the \robust\ are similar to those experienced by other software engineers.
    &
    This calls for prioritization of research and innovation. Consider whether it is more beneficial to focus on general software technology or on specific tools aiming at narrower problems.
    \looseness -1
    \\

    Many of the general programming mistakes within \robust\ could, in theory, be caught by existing, general-purpose QA tools and practices.
    &
    More action research combining software engineering researchers and robot programmers is needed to better understand this gap, to increase awareness of QA methods \& tools within robotics, and to make software engineering methods more accessible to programmers from non-computing backgrounds.
    \looseness -1
    \\

    Detecting system-level failures relies \emph{heavily} on human oracles;
    Specifying and monitoring intended behaviour is fundamentally difficult.
    &
    \robust\ provides 115 examples of such bugs that can be used to drive research into oracle specification and inference and automated testing for robotics software more broadly.
    \\

    Many bugs are detected during systems integration.
    \robust\ bugs were reported against software components, thus, in principle, they should be detectable in earlier life-cycle stages.
    \looseness -1
    &
    \robust provides a dataset to experimentally assess whether, indeed, unit testing, fuzzing, combinatorial testing, or other automated techniques could have detected these faults early, and if so, at what cost.
    \\

    Failures are complex but fixes are often simple. However, they are heavily domain- and project-specific.
    &
    We recommend developing generic methods for building lightweight program analysis and program fixing tools, in the spirit of language workbenches (lightweight tools for developing languages).

\end {tabularx}

\end {table}% __]

\paragraph{Building high-quality reusable robotics software components is difficult.}

ROS's modular design, rich package ecosystem, and spatially decoupled run-time architecture allow its users to quickly and easily build diverse robots using off-the-shelf hardware and software components without being an expert in all areas of robotics\,\citep {Kolak20}. \emph {However, those same characteristics make it difficult or impossible to anticipate incorrect or unintended interactions between that ROS module and its application, firmware, operating system, and hardware context.} Sources of variability include: hardware components, models, and firmware; operating system distributions; language and compiler versions; ROS parameters and launch arguments; and package versions. Even if the source code for a ROS node remains unchanged, its environment will continue to evolve, creating opportunity for unexpected failures.

Exhaustively testing all possible variations to find potential failures is both technically challenging and prohibitively expensive. Instead, the risk of an unexpected failure due to unexpected interactions can be mitigated by identifying, minimizing, and testing sources of variability within a process of continuous integration and deployment. Most software-related sources of variability (e.g., OS distributions, language and compiler versions, package versions) can be controlled effectively by using containers (e.g., Docker) to package, distribute, and deploy specific versions of individual nodes. Indeed, the use of Docker containers has gained popularity within the ROS community\,\citep {discourse-docker}. Hardware-related sources of variability (e.g., differing models, revisions, firmware) are harder to minimize and must be tested via the slow, expensive, and complicated process of software--hardware (SW/HW) integration testing\,\citep {AfzalICST20}. This testing is often performed in an ad-hoc manner by taking a joystick and manually running the robot through several scenarios, chosen based on intuition. However, more sophisticated setups have been used to perform continuous integration of hardware and software by deploying new software releases to a fleet of robots within a controlled testing facility outfitted extensively with a variety of sensors\,\citep {Henning2016}.

\emph {Given the inherent costs and associated challenges of dynamically detecting errors due to misconfiguration, there is a need for tools that can reliably detect the presence of such errors ahead of testing and deployment.}
One approach is to require or allow additional specification from users~(\citealt{aadl-ros,armassist,ros-skills}). Indeed, ROS 2 is moving in this direction. For example, ROS 2 requires that parameters be declared and allows users to specify expected types and value ranges that are checked at run-time~\citep{ros2nodeapi,ros2paramtutorial,ros2designparams}. Care needs to be taken however, as increasing the specification burden upon users can be both heavyweight (i.e., writing programs takes longer) and constraining (i.e., not all programs are covered). As an alternative to relying on increased specification, a number of papers have proposed the greater use of inference~\citep{Witte18,haros,haros-tool,haros-extraction,rosdiscover,rosinfer}. For example, tools such as HAROS and ROSDiscover combine the coarse-grained architectural information provided in ROS configuration files with static analysis of source code to approximate run-time architectures. Whether inference tools can be sufficiently powerful to produce accurate results and eliminate the specification burden remains an open question. Instead, the most effective approach may be in building inference techniques that exploit lightweight specification that is quick and easy for developers to provide, such as that required by ROS 2 parameters.
\looseness -1

\paragraph{ROS developers make the same mistakes as other developers.}

Most ROS software is written in C++ and Python. Naturally, it inherits the same set of general programming mistakes that are observed in those languages. Given the potentially catastrophic consequences of failure within the domain of robotic systems, it is vital that we reduce the space of possible failures by eliminating the possibility for common programmer mistakes (e.g., memory management, unhandled exceptions, missing synchronization).
This may be achieved either through the use of safe languages (e.g., Rust~\sqcitep{rust}, Erlang~\sqcitep{erlang}),
the introduction of new safety features and programming abstractions into existing languages
(e.g., RxROS~\sqcitep{rxros}),
or the development of easy-to-use analysis tools that reliably detect certain classes of
error~(e.g., PhrikyUnits~\sqcitep{Phriky}, Phys~\sqcitep{Phys}, PhysFrame~\sqcitep{physframe}).
Solutions need to be accessible to ROS's broad demographic of users, without compromising run-time performance or entirely sacrificing the lightweight, prototypical aspects of ROS. Advanced computer science knowledge should not be assumed \citep{Alami18}.
\emph {It is a crucial finding of this study that, in fact, most software development problems faced by robotics developers, as documented in issues and pull requests in \textnormal {\robust}, are similar to those experienced by other software engineers.  The robotics-specific issues do not dominate the landscape.}
\looseness -1

\paragraph{Many problems are in classes addressed by existing tools.}

Just as ROS developers make the same mistakes as other developers, \emph {we observe that many of those mistakes could be caught by existing, general-purpose QA tools and practices.} For example, incorrect or missing build and run-time dependencies can be quickly detected by automatically building and fuzzing as part of a continuous integration pipeline. Certain runtime errors bugs caused by, e.g., out-of-bounds memory access, data races, and resource mismanagement, may also be detected statically via analysis tools (e.g., Infer~\sqcitep{infer}, Footpatch~\sqcitep{footpatch}) or at runtime via monitoring instrumentation (e.g., AddressSanitizer~\sqcitep{asan}, MemorySanitizer~\sqcitep{msan}, ThreadSanitizer~\sqcitep{tsan}). \emph {A key challenge lies in identifying and eliminating the barriers to adoption for these tools and techniques within the open source robotics community.}

\paragraph{Bug detection relies heavily on human-in-the-loop testing.}

Many software failures in the dataset (\numbugssysnone{}\,of\,\numbugs{}) have no noticeable effect at the system level.
As many as~\numbuildtimebugs{} of the~\numbugssysnone{} occur at build time, are easily detected, and, by definition, have no effect on system behavior.
The remaining~\numbugsysnonebutnotbuild{} failures, however, are difficult to detect, lacking immediately noticeable effects on system-level behavior.
\looseness -1

The \numbugssys{} bugs that lead to externally observable symptoms vary considerably in terms consequences, ranging from mild annoyance to potential catastrophe depending on the operational environment. \emph{A common theme among these bugs is that their detection relies \emph{heavily} on human oracles: It is hard to specify expected behaviour at the system level,  especially in terms of observable physical effects, and it is harder still to automatically observe and analyze such behaviour.} The 115 bugs mentioned above are by themselves a research agenda for the testing community.

Indeed, the challenges of writing tests are reflected in \robust: fewer than 10\% of relevant bug fixes (\numfixedbugswithtests of \numruntimeandstartuptimebugs) are accompanied by a test case (\cref{sec:method}).
Other studies have made a similar observation. In a sociocultural study of the ROS community, \citet{Alami18} found that the ROS community values the creation of new features and functionality, QA tasks are perceived as difficult and time consuming, are often undervalued and neglected.
In a series of semi-structured interviews of robotics practitioners, \cite{AfzalICST20} find that field testing is the predominant means of QA within robotics, and that the challenges prevent or dissuade developers from writing automated tests.

\emph {Given the importance and difficulty of writing tests for robotics software, there is an urgent and growing need for new tools, techniques, and infrastructure to allow developers to easily and effectively test their robotics software.} To that end, researchers have proposed test input generation techniques for individual components (e.g., \citealt{haros-testing}) and entire systems (e.g., \citealt{rvfuzzer}),
methods for inferring and monitoring intended robot behavior (e.g.,~\citealt{artinali,arsi,Inoue2017,Mithra}), and domain-specific languages for describing simulated testing environments (e.g.,~\citealt{scenic,paracosm,kluck2018,gzscenic}).
\looseness -1

From a software quality point of view, employing a variety of techniques (possibly in a dependability case argument) is likely the best approach.
In general, it might be easier to simply detect the underlying software faults with automatic or semi-automatic tools, than it is to detect the failure at the system level.
At the very least, all the bugs collected in \robust\ have been reported to issue trackers of software components, which indicates that a regression tests for them might be possible.
These further reinforces our view that failures should be shifted \emph{to the left} as far as possible, closer to build time, and ideally contained in single components.
\looseness -1

\paragraph{Failures are complex; fixes are often simple.}

While failures can be complex, unpredictable, and affect multiple components in several languages, over 90\% of bug fixes are limited to a single language, almost two thirds of bugs are fixed in a single file, and most bug fixes are relatively small in terms of their number of lines added, modified, and deleted (\cref{sec:method}). Program repair techniques are most effective when the necessary syntactic changes are both small (i.e., require changes to few lines of code) and isolated to a particular region of the program (e.g., a single file or method). \emph {Thus, our findings are encouraging for existing APR techniques whose success is likely to be bound by the lack of sufficient tests rather than the complexity of the necessary repairs.}

That most bugs are fixed in a single language is encouraging for development of analysis and automated repair tools, which are typically limited to a single language. However, many bug fixes occur in domain-specific languages or dialects that lack analysis tools, or else involve interactions with external components that are difficult to deal with using existing techniques.
\emph {While research on domain-specific languages resulted in many efficient and generic techniques for their implementation \citep {lammel:2018,wasowski.berger:2023}, static analysis tools and testing methods for domain-specific models have hardly received the same attention, even though, as our data indicates, they would be potentially useful in robotics software engineering practice.}
\looseness -1

We list ROS-specific analysis tools and techniques that help to tackle some of these challenges. Techniques have been proposed for both statically and dynamically determining and verifying interactions (e.g., over topics and services) between ROS components~\citep {Witte18,haros-extraction}. \citet {Purandare12} implement an analysis that helps to explain changes in conditional message flows (i.e., the conditions under which messages are published) between program versions to aid in debugging. \citet {DBLP:conf/icse/Fischer-Nielsen20} provide an approach for automatically identifying and fixing missing ROS package dependencies. HAROS~\citep {haros} provides a graphical front-end for presenting and integrating the results of various analyses on ROS systems.
\looseness -1

\section{Benchmark Infrastructure}%
\label{sec:infrastructure}

To allow our dataset to be used to assess new QA tools and techniques, it is vital that we provide access to accurate recreations of the build and run-time environments in which each of the identified bugs took place. Reproducing historical software environments is never trivial, but ROS, its tools, and the operating systems it runs on proved to be an exceptionally volatile mix in this project. The interdependence between tools and the required versions of software libraries complicated building up the infrastructure needed to allow restoration of the bugs in \robust. While we recognise this infrastructure is specific to the context of ROS, the challenges it addresses are not specific to ROS, and would be experienced when restoring historical snapshots in any ecosystem depending on heterogeneous, independently evolving components.

\paragraph{Identification and Retrieval of Historical Source Code.}

A historical bug is embedded in source code with many static and dynamic dependencies.
In order to build the system with the bug, one needs the exact version of the buggy source code with all its build and run-time dependencies as they were at the point in time where the bug was identified---a \emph{contextual restoration snapshot}.
The following properties of historical ROS bugs necessitate the recreation of a snapshot of more than just the buggy component:
\looseness -1

\begin {enumerate}

  \item The bug has likely been fixed since it was reported which means it can no longer be found in up-to-date versions of the source code.

  \item The code containing the bug, along with its dependencies and dependants, have evolved, which precludes compilation of that code on current systems.

  \item ROS is an evolving platform, which prevents reproduction with new versions; even properties other than source-code-level compatibility change, e.g., run-time behavior of components and semantics of message exchanges.\looseness -1

\end {enumerate}
The possibility of creating a contextual restoration snapshot depends entirely on our ability to identify the exact version of the source code which caused the bug.
For \robust, this information has been extracted from issue reports, pull requests, and by inspection of the revision log of the version control systems used by the authors and maintainers of the analyzed components.
For each bug, we establish the contextual reproduction snapshot as either: (a) the \emph {parent commit} of the bug-fixing commit, if the bug has been fixed, or (b) the last commit \emph {preceding} the date and time at which the bug was reported. The snapshot does not contain the fix, but the buggy code.
\looseness -1

\begin{figure}[t]% [__

  \centering
  \includegraphics[
    width = .6 \textwidth,
    clip,
    trim = 14mm 38mm 14mm 38mm
  ]{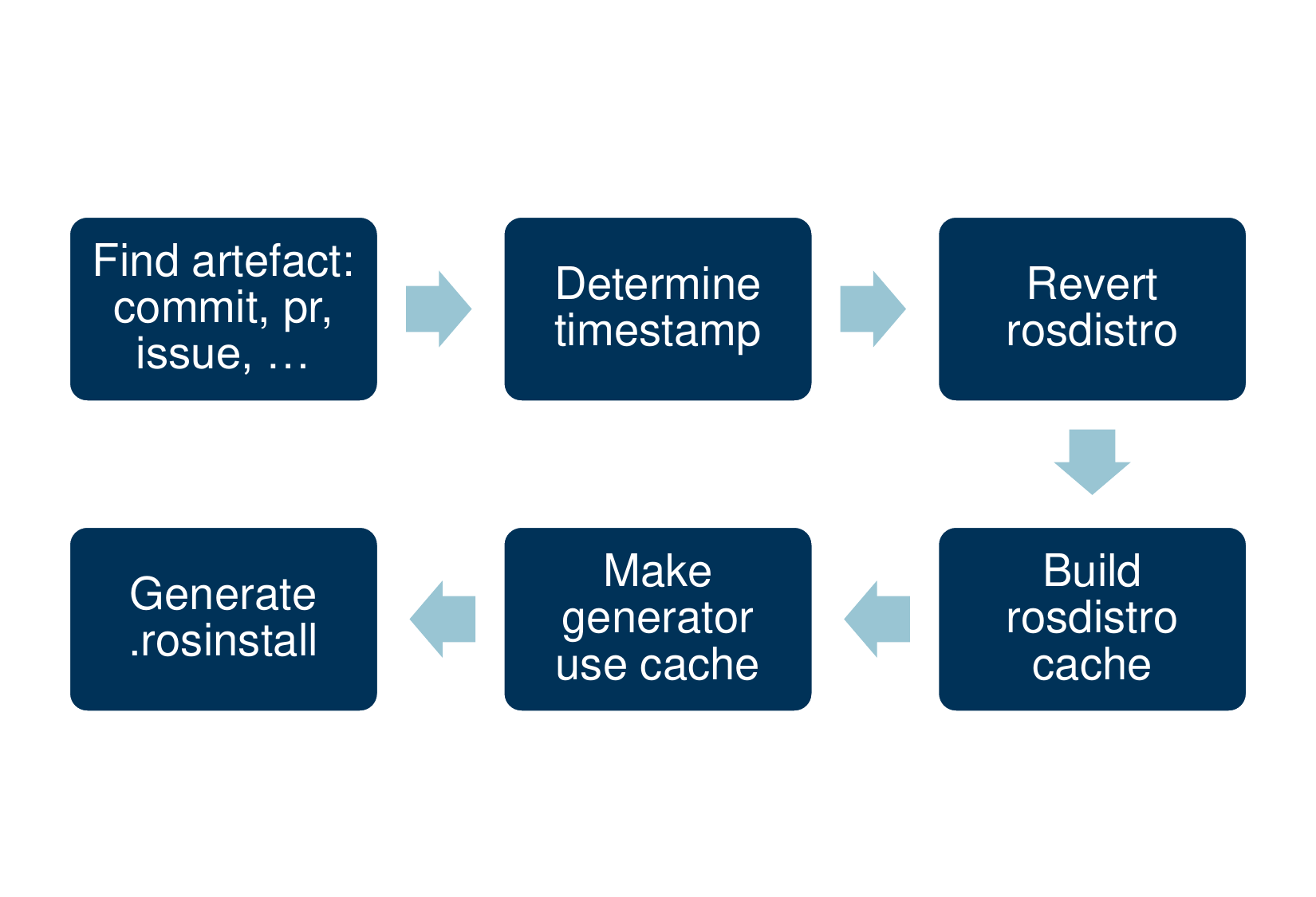}

  \caption{Diagram of the rosinstall generator time machine workflow.}%
  \label{fig:ritm_diagram}

\end{figure}% __]

\paragraph{Recreation of the ROS Distribution.}%
\label{sec:infra-time-mach}

\begin{figure}[!t]% [__
\centering
\begin{lstlisting}[language=yaml]
kobuki_core:
  doc:
    type: git
    url: https://github.com/yujinrobot/kobuki_core.git
    version: kinetic
  release:
    packages:
    - kobuki_core
    - kobuki_dock_drive
    - kobuki_driver
    - kobuki_ftdi
    tags:
      release: release/kinetic/{package}/{version}
    url: https://github.com/yujinrobot-release/kobuki_core-release.git
    version: 0.7.8-1
  source:
    type: git
    url: https://github.com/yujinrobot/kobuki_core.git
    version: kinetic
  status: maintained
\end{lstlisting}
\caption{\texttt{ros/rosdistro} entry for Kobuki, showing several released packages
  and the location of the repository containing the source code (ROS Kinetic,
  \texttt{ros/rosdistro@cdfe60d2}).%
}%
\label{fig:rosdistro_snippet}
\end{figure}% __]

\Cref{fig:ritm_diagram} shows a visual representation of our approach.
Once we have identified the contextual reproduction snapshot and its associated time, we reproduce the ROS distribution as it was ``back then,'' by computing the transitive closure of all dependencies of the subject package.
For a typical bug, this set can range from three to a hundred packages, both ROS and standard OS packages (called \emph{system dependencies} in ROS).\@
We gratefully make use of the index of all \emph{released packages} in a distribution, centrally maintained in a GitHub repository known as \texttt{rosdistro}.\footnote{\url{https://github.com/ros/rosdistro}}
A file called \texttt {distribution.yaml} in this repository lists all packages released into a specific ROS distribution including their source repositories and the commit hashes and branches used for a particular release.
\Cref{fig:rosdistro_snippet} shows an entry for the \texttt{kobuki\_core} package for a particular ROS distribution (Kinetic) at a specific point in time.
We revert the \texttt{rosdistro} database back to the right point in time and then use a ROS tool called the \texttt{rosinstall\_generator} to compute the previously mentioned closure of dependencies.
The result of this, a \texttt{rosinstall} file, describes the locations and versions of all required dependencies along with a pointer to the repository containing the reproduction snapshot of the package which contains the bug.
Figure\,\ref{fig:rosinstall_file_45ee84a} shows an excerpt from such a \texttt{rosinstall} file for the \bugkobuki{45ee84a} bug.
\looseness -1

\begin{figure}[!t]% [__
\centering
\begin{lstlisting}[language=yaml]
- tar:
    local-name: catkin
    uri: https://github.com/ros-gbp/catkin-release/.../catkin/0.6.11-0.tar.gz
    version: catkin-release-release-indigo-catkin-0.6.11-0
- tar:
    local-name: ecl_command_line
    uri: https://github.com/yujinrobot-release/.../ecl_command_line/0.61.0-0.tar.gz
    version: ecl_core-release-release-indigo-ecl_command_line-0.61.0-0
- tar:
    local-name: ecl_license
    uri: https://github.com/yujinrobot-release/.../ecl_license/0.61.0-0.tar.gz
    version: ecl_tools-release-release-indigo-ecl_license-0.61.0-0
\end{lstlisting}
\caption{\texttt{rosinstall} file listing the dependencies of \texttt{kobuki\_ftdi}
in the ROS Indigo distribution (bug: \bugkobuki{45ee84a}). Note: \texttt{uri}
entries have been shortened.%
}%
\label{fig:rosinstall_file_45ee84a}
\end{figure}% __]

We then use this file to clone all the required repositories at the right commit and build them using their original (CMake) build scripts.
(Of course, for non-ROS packages this procedure will be slightly different as they may use a different build-system, however the process largely remains the same.)

Admittedly, this restoration procedure does not capture the versions of all dependencies as they were exactly at the snapshot time on the particular computer when and where the bug was identified and potentially fixed, but rather, it selects the versions that were \emph{released} at that point in time.
We assumed this to be sufficient for most cases as developers typically
work with the latest available releases of packages.
Of course, this procedure also will not capture cases where a developer had either a really old, really new, or unreleased version of a particular dependency installed and the bug depends intimately on that particular version.
For the bugs described in \robust, however, we determine that this is seldom the case, and where it was discovered this played a role in reproduction of the bug, manual changes to the output of the \texttt{rosinstall\_generator} were made to account for this.

Reproducing the historical version of a ROS distribution is one of the hardest steps in the process.
The existence of tools in ROS, such as the generator and the dependency modelling and introspection tools, has facilitated this greatly.
In projects where seeding of development environments is less structured and automated or is not using explicit package management and metadata, along with the corresponding tools, this would have been much harder.

\paragraph{Restoration of the Non-ROS Environment.}

Even if we have successfully identified and copied the historical versions of the source code and computed and cloned all the correct dependencies of packages, compilation often fails on machines with up-to-date build environments.
This is caused by the fact that operating systems, system libraries, run-time environments, compilers, interpreters and build tools are continuously evolving and the subject package may depend on particular versions of any of these.
Such dependencies may even be only implicitly embedded in the package, as explicitly stating versions of system dependencies is not often done in ROS packages.
Furthermore, dependencies may become unavailable on newer systems.
For example, code repositories could disappear, operating system distros could become unavailable, or the compiler or other toolchain elements could have been discontinued entirely.

In the period covered by \robust, between 2010 and 2018, Canonical released 10 different Ubuntu versions (and still supported three older LTS releases, GCC saw 36 releases (including patch releases), Boost (a main dependency of ROS and thus all ROS packages) released 20 versions and Python supported a range of both 2.x and 3.x versions at the same time, spanning from version 2.6 on Ubuntu Lucid to version 3.5 on Ubuntu Bionic.
ROS itself saw 12 releases in the same period, with multiple ROS versions supporting multiple Ubuntu versions (\cref{tab:ubuntu_vs_deps_vs_ros}).
\looseness -1

\begin{table}[t]% [__
\centering
\small
\begin{tabular}{llrrll}
\textbf{Ubuntu} & \textbf{ROS} & \textbf{GCC} & \textbf{Boost} & \textbf{Python 2} & \textbf{Python 3} \\
\midrule
Lucid   (10.04) & C, D, E, F & 4.4 & 1.40 & 2.6.5  & 3.1.2 \\
Precise (12.04) & F, G, H    & 4.6 & 1.48 & 2.7.3  & 3.2.3 \\
Trusty  (14.04) & I, J       & 4.8 & 1.54 & 2.7.5  & 3.4 \\
Xenial  (16.04) & K, L       & 5.3 & 1.58 & 2.7.12 & 3.5.1 \\
Bionic  (18.04) & M          & 7.4 & 1.65 & 2.7.15 & 3.6.7 \\
Focal   (20.04) & N          & 9.3 & 1.71 & 2.7.17 & 3.8.2 \\
\bottomrule
\end{tabular}
\caption{Versions of ROS supported on various Ubuntu versions (Long Term Support releases only) and the versions of core dependencies of ROS on those platforms. Ubuntu versions listed by their codenames. ROS versions listed only by the first character of the codename: \emph{C-Turtle}, \emph{Diamondback}, \emph{Electric}, \emph{Fuerte}, \emph{Groovy}, \emph{Hydro}, \emph{Indigo}, \emph{Jade}, \emph{Kinetic}, \emph{Lunar}, \emph{Melodic} and \emph{Noetic} respectively.}
\label{tab:ubuntu_vs_deps_vs_ros}
\end{table}% __]

We use containerization to manage the historical contexts required by the bugs in \robust.
Each entry is accompanied by a Docker image containing the appropriate historical code snapshot accompanied by the correct versions of tools and the right build and run-time environment needed for reproduction.
Containers also allow for relatively easy and efficient distribution of pre-built environments, which reduce the time it takes researchers to start working with these bugs, as sessions can be started after a download of the pre-built context, instead of having to wait for one to be built on demand.
\looseness -1

To avoid having to handcraft images for each bug in \robust, we make use of \emph{BugZoo}~\citep{timperley2018bugzoo}, which simplifies the specification of parameters and build file generation.
BugZoo takes care of building and launching Docker containers for specific bugs by using information from the \robust bug description file combined with the \texttt{rosinstall} files.
As many bugs share the same basic container structure, a template \texttt{Dockerfile} is used by BugZoo to generate the recipes for specific bugs.

\paragraph{Extensibility.}

By following the process described in Section~\ref{sec:subjects} more bugs could be added, and not just for the packages already analysed by the authors.
The same procedure could be used for any ROS package, the only requirement being that the package has been released at some point in its lifetime.
This is due to limitations in the tooling used to compute the transitive closure of dependencies for a specific package, which depends on \texttt{rosdistro} for information about those dependencies and the state of other ROS packages part of the historical snapshot.
\looseness -1

Researchers from both the software repair and the ROS community could submit new bugs for inclusion into the database by submitting pull requests against the \href{https://github.com/robust-rosin/robust}{\texttt{robust}} GitHub repository.

\paragraph{Durability.}\label{sec:infra:durability}

While using container technology allows us to safeguard reproduction snapshots against decay and becoming incompatible with ever evolving OS and run-time environments, by itself, this technology does not solve the problem of persistence of the dataset.
Container technology itself will most likely evolve, and the implementation which today enjoys the greatest popularity may not exist in the future, which would reduce all our containers to static data archives instead of packaged, executable environments.
While we cannot prevent this from happening, we have employed several different tactics to mitigate the impact of these types of bitrot, to the point where recreation of the snapshot itself should at least be possible if direct reuse of the containers themselves turns out to be difficult.

First, the process of container creation is as much as possible captured in code, and documented in natural language.
These plain-text artifacts are all committed to an open-source Git repository.

Second, all technology used for the creation of the current containers is open-source, without relying on any proprietary tools.
Proprietary tools have a clear disadvantage when it comes to reproducibility:
without source code available, rebuilding them on future platforms will be impossible and run-time support limited to platforms which are compatible today.
By comparison, open-source tools, while perhaps also not immediately usable, offer at least the possibility of fixing any limitations which might prevent them from being used in newer run-time environments.

Third, even though the source repositories which host the subject systems are all open-source and available online, we create forks of those repositories and host them in the \href{https://github.com/robust-rosin}{\texttt{robust-rosin}} organization on GitHub.
Those forks are then used to build Docker container images.
This organization is under our control, and forking (i.e., copying with preservation of provenance) creates a duplicate of the forked repository which prevents changes to the source repository from tampering with the code history or structure.

Finally, every container includes a copy of the source code repositories which make up the reproduction snapshot.
These repositories were identified using the process and tools described above and include both the subject package (containing the buggy code) as well as its (transitive) dependencies.
Should a container no longer be functioning, these sources could be extracted from it and used to repopulate a new reproduction environment.

\section{Related Work}\label{sec:related}

\paragraph{Bug Datasets.}

A variety of bug datasets have been proposed as benchmarks for evaluating
fault localization, fuzzing, test prioritization, bug diagnosis, and program repair
techniques. Datasets have been created for a variety of languages and platforms
including JVM-based languages~\citep{defexts}, Java~\citep{defects4j,bears,bugsjar,quixbugs,bugswarm},
Android Java~\citep{droixbench}, NodeJS~\citep{bugsjs}, Python~\citep{refactory,quixbugs,bugswarm,bugsinpy},
and C~\citep{corebench,manybugs,dbgbench,codeflaws,sir}.

% source code
Some of the buggy programs within these datasets are provided exclusively as source code,
and do not come with a self-contained environment for accurately reproducing
their behavior. Over time, this can lead to programs that are unusable or that
produce different results, often due to unstated dependencies (e.g., libraries and compilers).
% VMs
Other datasets, such as ManyBugs and IntroClass~\citep{manybugs}, are provided
as a single virtual machine (VM) image.
While VMs provide a stable environment for conducting experiments,
they often lack transparency (i.e., it is unclear how they were produced),
make it difficult to use new software and tools inside old VMs,
and have considerable performance overheads,
making them unsuitable for time-sensitive and resource-intensive systems such
as those used in robotics.

Most recently, datasets have moved toward using Docker containers as the preferred
means of distribution.
BugSwarm is a relatively new benchmark that currently consists of over 3000 bug fixes in
open-source Java and Python projects. BugSwarm is continually populated
by mining GitHub and TravisCI to find regression fixes.
In our dataset, we observe that tests are rarely provided with bug fixes and that tests are lacking, generally, indicating that BugSwarm's approach would be
unable to find most of the run-time bugs of our dataset.

We build on top of BugZoo, which packages bugs as individual Docker container images
and a set of machine-readable instructions (e.g., for building and and testing the
buggy program), to allow reproducible interactions.
We significantly advance the state of the art in bug reproduction through
the introduction of our \enquote{time machine,}
which obtains a historically accurate context for each bug,
and a generic Docker
setup that can be used to recreate historical versions of arbitrary ROS packages.
Our machinery and methods can be reused by others to study a greater portion
of software within the ROS ecosystem.
\looseness = -1

\paragraph{Bug Studies.}

Our methodology for analyzing and documenting bugs is inspired by
Abal et al.~\citep{varlinuxbugs,varlinuxbugs-journal} qualitative study
of variability bugs in the Linux kernel. Their dataset
provides a natural language description, execution trace, CWE classification,
location,
presence condition, and traceability information (e.g., repository URL,
commit hash, associated issues) for each bug.
Our approach also includes these elements, where relevant, and goes further
by incorporating a qualitative analysis of faults and failures, and providing
historically accurate Docker images for each bug.

Numerous empirical studies have investigated the characteristics of
bugs in various contexts and of various nature: bugs in test code~\citep{Vahabzadeh15},
concurrency bugs~\citep{Fonseca10,Wang17,Asadollah17},
configuration bugs~\citep{Yin11}
performance-related issues~\citep{Jin12,Han16,Yang18},
how developers diagnose, debug, and fix bugs~\citep{dbgbench}. Others have
examined faults and failures within various domains including
machine learning~\citep{Thung12,Islam19,Humbatova20},
blockchain~\citep{Wan17},
Android~\citep{Bhattacharya13},
and operating systems~\citep{varlinuxbugs,Chou01}.
In this study, we construct and analyze a dataset of bugs in popular ROS software,
to provide (a) insight into the nature of software bugs in open-source robotics
and (b) a testbed for building and evaluating QA tools and techniques for ROS.

A smaller number of papers have examined faults and failures within robotics systems software.
\cite{Steinbauer12} conducts a quantitative survey of research groups that participated
in RoboCup, an annual international robotics competition, to learn more about
software, hardware, algorithmic, and human-interaction faults in robotic systems.
\cite{Grottke10} identify and characterize 520 software faults in the on-board
software for 18 JPL/NASA space missions
in terms of their
ease of reproducibility by studying how they are triggered and
their resulting error propagation.
\cite{ardubugs} perform an empirical study of 228 historical bugs in ArduPilot~\citep{ardupilot},
a popular autopilot for a variety of autonomous vehicles, to determine
whether those bugs are reproducible in simulation.
In comparison to those studies, we restrict out attention to software bugs and
curate, examine, and document individual bugs across a diversity of ROS software.

\paragraph{Comparison to Existing Taxonomies}

\begin{table}[t]
\small
\begin{minipage}{.49\linewidth}
\centering
\begin{tabular}{c|c}
\toprule
\strut \textbf{\citeauthor{Zampetti22}} & \textbf{Ours} \\
\midrule
\strut Algorithm     & \fullcirc\xspace Programming \\
\strut Config        & \fullcirc\xspace BDO, Config \\
\strut Data          & \fullcirc\xspace Programming \\
\strut Documentation & \emptycirc\xspace \\
\strut Hardware      & \halfcirc\xspace Systems \\
\strut Interface     & \fullcirc\xspace Programming \\
\strut Network       & \emptycirc \\
\strut ~\\
\bottomrule
\end{tabular}
\end{minipage}%
\begin{minipage}{.49\linewidth}
\centering
\begin{tabular}{c|c}
\toprule
\strut \textbf{\citeauthor{Wang21}} & \textbf{Ours} \\
\midrule
\strut Limit          & \fullcirc\xspace Config, Models \\
\strut Math           & \fullcirc\xspace Programming \\
\strut Inconsistency  & \fullcirc\xspace Models, Systems \\
\strut Priority       & \fullcirc\xspace Concurrency \\
\strut Parameter      & \fullcirc\xspace Config \\
\strut H/W Support    & \fullcirc\xspace Systems \\
\strut Correction     & \fullcirc\xspace Models, Programming \\
\strut Initialization & \fullcirc\xspace Programming \\
\bottomrule
\end{tabular}
\end{minipage}%
\caption{A comparison of the root causes presented in \cite{Wang21} and \cite{Zampetti22} against our fault taxonomy.
\fullcirc\xspace indicates that all bugs in their category fall into one or more of our categories.
\halfcirc\xspace indicates that some of the bugs in their category fall into one or more of our categories.
\emptycirc\xspace indicates that none of the bugs in their category are represented by any of our categories.
}
\label{tab:fault-taxonomy-comparison}
\end{table}

\begin{table}[t]
\small
\centering
\begin{tabular}{c|p{0.33\textwidth}p{0.33\textwidth}}
\toprule
\textbf{Ours} & \textbf{\citeauthor{Zampetti22}} & \textbf{\citeauthor{Wang21}} \\
\midrule
BDO & \fullcirc\xspace Config & \emptycirc \\
Config & \fullcirc\xspace Config & \halfcirc\xspace Limit, Parameter  \\
Concurrency & \fullcirc\xspace Concurrency & \halfcirc\xspace Priority \\
Evolution & \emptycirc & \emptycirc \\
Programming & \halfcirc\xspace Algorithm, Data, Interface & \halfcirc\xspace Correction, Initialization, Limit, Math \\
Models & \emptycirc & \halfcirc\xspace Correction \\
Systems & \halfcirc\xspace Hardware & \halfcirc\xspace H/W Support \\
\bottomrule
\end{tabular}
\caption{A comparison of the coverage of existing taxonomies vs. our own.
\fullcirc\xspace indicates that our label is fully covered by a set of their labels.
\halfcirc\xspace indicates partial coverage,
and \emptycirc\xspace indicates no coverage.}
\label{tab:compared-vs-ours}
\end{table}

Below, we present a comparison of our the results of our study against those obtained from other, recent studies of bugs in similar domains:
\cite{Garcia20} study 499 bugs in two, popular autonomous vehicle software systems, Apollo~\citep{Apollo} and Autoware~\citep{Autoware};
\cite{Wang21} examine 569 bugs across two well-known UAV software platforms: PX4~\citep{px4} and ArduPilot~\citep{ardupilot};
and \cite{Zampetti22} analyze 655 bugs across 14 projects related to cyberphysical systems, including Autoware, ArduPilot, and PX4.
All three approaches construct a dataset by identifying issues~\citep{Wang21,Zampetti22} or merged pull requests~\citep{Garcia20} in GitHub projects.

In terms of faults, \citeauthor{Garcia20} use an open coding to adapt an existing taxonomy of root causes~\citeauthor{Thung12}
and \citeauthor{Zampetti22} follow a similar approach to extend \citeauthor{Garcia20}'s taxonomy.
After identifying 569 bugs in their subject systems, \citeauthor{Wang21} use open coding to devise a taxonomy for 168 of those bugs (29.5\%) that they determine to be UAV-specific.%
\footnote{Note that they make a similar observation to our own study (see Section~\ref{sec:findings}) that the majority of bugs that they studied were not specific to UAVs.}

\Cref{tab:fault-taxonomy-comparison} provides a mapping from those existing taxonomies to our own.\footnote{We do not compare directly to \cite{Garcia20} since \cite{Zampetti22}'s taxonomy subsumes it.}
Despite being specific to UAVs, all of the bugs within each of \citeauthor{Wang21}'s taxonomy fit into one or more of our categories.
Most of \citeauthor{Zampetti22}'s taxonomy can be mapped onto our own with the exception of
\enquote{Documentation}, which we consider to be out of scope,
\enquote{Network}, which we consider as a software-level failure caused by an underlying fault (e.g., general programming),
and parts of their \enquote{Hardware} category (e.g., energy, hardware failure) since we restrict our study to software issues.

\Cref{tab:compared-vs-ours} provides a mapping from our taxonomy to theirs.
None of the taxonomies identify \enquote{Evolution} as a fault, nor do they fully capture our \enquote{Models} category.
While ArduPilot, Autoware and PX4 are mostly self contained,
our subjects exhibit the norms of ROS development and represent either (a) reusable components that are used to build a larger system (e.g., Geometry2, MAVROS) or (b) robot systems that are built from third-party components (e.g., Care-o-Bot, Kobuki, TurtleBot).
ROS's component-based architecture and emphasis on reusability with minimal assumptions makes ROS software more prone to \enquote{Evolution} issues.
Similarly, as components are often designed to be agnostic to the specifics of any particular robot (e.g., hardware, environment, application), inconsistencies are relatively easy to introduce but difficult to check.

In terms of failures, neither \citeauthor{Wang21} nor \citeauthor{Zampetti22} study the failures associated with the bugs within their datasets.
\citeauthor{Garcia20} derive a new taxonomy of bug symptoms (i.e., failures) through of a combination of examining previous taxonomies of bugs in machine learning~\citep{Thung12}, deep learning~\citep{Zhang18,Islam19}, and numerical software~\citep{DiFranco17} and following an open coding approach on their dataset.
They make no explicit distinction between software and system-level failures, but both levels appear within their taxonomy.
In terms of software-level failures, our \enquote{Performance} and \enquote{Network} categories are not covered.
\citeauthor{Garcia20}'s system-level failures are focused on the outputs of AV subsystems (e.g., incorrect trajectory prediction, localization, lane positioning and navigation).
In constrast, our system-level failures provide a different perspective and are focused on \emph{how} the failure manifests (e.g., loss of functionality, unresponsiveness, degraded functionality) as opposed to \emph{where} it manifests.

In comparison to the prior studies, our dataset is application and architecture agnostic, and we examine a diverse set of reusable libraries, components, and platforms that are used to build robot applications.
Our results highlight interesting differences in the bugs that affect ROS software and provide further support for the findings of previous studies by performing a conceptual replication (i.e., devising a taxonomy \emph{de novo}).
Crucially, we uniquely provide detailed reports for every bug within our dataset (c.f. a URL and a list of labels), along with Docker images and associated tooling, all of which allow researchers to study robotics bugs in more depth and to use our dataset as a benchmark for assessing QA techniques.

\section{Conclusion}\label{sec:conclusion}

In this paper, we presented \robust, a study and accompanying dataset of 221 bugs in Robot Operating System software. We systematically collected, documented, and analyzed bugs across 7 popular ROS projects, and produced Docker images that allow researchers building QA tools and techniques for robotics to interact those bugs. We classified faults and failures within a taxonomy we constructed for this purpose, based on a qualitative analysis of our dataset, highlighted findings of particular interest to the software engineering research community, and discussed the ramifications of our results. The \robust\xspace dataset and more information about the project can be found at our companion website: \url{https://github.com/robust-rosin/robust}.

\begin {acknowledgements} We thank Claus Brabrand, Zhoulai Fu, Jon Azpiazu, Jon Tjerngren, Jonathan Hechtbauer, Jam Marcos Hernandez, and Claire Le Goues for discussions about the study design, motivation, and for contributing bug analyses or technical knowledge in the process.
This work was partially supported by the European Union’s Horizon 2020 research and innovation programme, grant agreement No. 73228 (ROSIN), Marie Skłodowska-Curie Actions grant agreement No 956200 (REMARO), by DARPA (\#A8750-16-2-0042), and by AFRL (\#FA8750-15-2-0075). The authors are grateful for their support. Any opinions, findings, or recommendations expressed are those of the authors and do not necessarily reflect those of the US Government or the European Union.
\end {acknowledgements}

\bibliographystyle {spbasic}
\bibliography {references}

\end{document}